\documentclass[twocolumn,twocolappendix,iop,apj]{emulateapj}
\usepackage{times}
\usepackage{comment}
\usepackage[T1]{fontenc}
\usepackage{amsmath,wasysym}
\usepackage{natbib}

\newcommand{\hi}{H\,{\sc i}}
\newcommand{\hii}{H\,{\sc ii}}
\newcommand{\fermilat}{{\em Fermi}--LAT}

\newcommand{\gray}{$\gamma$-ray}

\newcommand{\GP}{{\it GALPROP}}
\newcommand{\GG}{{\it GALGAS}}

\begin{document}

\title{The Three-Dimensional Spatial Distribution of Interstellar Gas
 in the Milky Way: Implications for Cosmic Rays and
 High-Energy Gamma-Ray Emissions}


 \author{Gu{\dh}laugur J{\'o}hannesson}
 \affiliation{Science Institute, University of Iceland, IS-107 Reykjavik, Iceland}
 \affiliation{Nordita, KTH Royal Institute of Technology and Stockholm University, Roslagstullsbacken 23, SE-106 91 Stockholm, Sweden}
\author{Troy A. Porter} \author{Igor V. Moskalenko}
\affiliation{W. W. Hansen Experimental Physics Laboratory and Kavli Institute for Particle Astrophysics and Cosmology, \\ Stanford University, Stanford, CA 94305, USA}


\begin{abstract}
  Direct measurements of cosmic ray (CR) species combined with observations of
  their associated \gray{} emissions can be used to constrain models of CR
  propagation, trace the structure of the Galaxy, and search for signatures of
  new physics. The spatial density distribution of the interstellar gas is a
  vital element for all these studies. So far models have employed the 2D
  cylindrically symmetric geometry, but their accuracy is well behind that of
  the available data. In this paper, 3D spatial density models for the neutral
  and molecular hydrogen are constructed based on empirical model fitting to
  gas line-survey data. The developed density models incorporate spiral arms
  and account for the warping of the disk, and the increasing gas scale height
  with radial distance from the Galactic center. They are employed together
  with the \GP{} CR propagation code to investigate how the new 3D gas models
  affect calculations of CR propagation and high-energy \gray{} intensity
  maps. The calculations made reveal non-trivial features that are directly
  related to the new gas models. The best-fit values for propagation model
  parameters employing 3D gas models are presented and they differ
  significantly from the values derived with the 2D gas density models that
  have been widely used. The combination of 3D CR and gas density models
  provide a more realistic basis for the interpretation of non-thermal
  emissions from the Galaxy.
\end{abstract}

\section{Introduction}

The distribution of the interstellar gas is a key ingredient to any
self-consistent model describing propagation of cosmic rays (CRs) and
generation of non-thermal interstellar emissions.
Propagation of CR species in the Galaxy and their interactions with
the interstellar gas and radiation field produce
changes in their composition and spectra due to
fragmentation, secondary particle production, and
energy losses \citep[see][for a review]{StrongEtAl:2007}.
Measurements of the
spectra and composition of CR species are used to constrain the most important
parameters of CR propagation models \citep[e.g.,][]{JohannessonEtAl:2016}.
Observations of the interstellar emissions generated through the production and
decay of neutral pions and inverse Compton scattering of CR electrons off
the interstellar radiation field (ISRF) provide a direct probe of the spatial densities and spectra
of CR protons, helium, and electrons in distant locations
\citep[e.g.,][]{AbdoEtAl:2009,AbdoEtAl:2010,AckermannEtAl:2011,AckermannEtAl:2012,AdeEtAl:2015,AjelloEtAl:2016}, far beyond the reach of
direct measurements.
The interpretation of these data requires well developed propagation models and a detailed
knowledge of the spatial distribution of the interstellar gas in the Milky Way.

The interstellar gas consists mostly of hydrogen and helium with a number
density ratio of approximately 10 to 1 \citep{Ferriere:2001}, while heavier
elements represent a negligible fraction of the total gas mass.
Depending on
its temperature, three forms of the hydrogen gas are distinguished: atomic (\hi),
molecular (H$_2$), and ionized (\hii) hydrogen, while helium remains
mostly neutral due to its much higher first ionization potential.
The \hi{} component is the
most massive, containing about 60\% of the mass while H$_2$ and \hii{} contain
25\% and 15\%, respectively \citep{Ferriere:2001}.
The spatial density
distribution of the three forms is also widely different.
The \hii{} component
is the most widespread with a large scale height perpendicular to the Galactic
plane of a few hundred pc near the solar system and a relatively low number
density.
The H$_2$ component on the other hand has a scale height of few tens
of pc near the solar system, is very clumpy, and contained mostly in high
density molecular clouds.
The distribution of the neutral \hi{} gas is
somewhat intermediate between those of the ionized and molecular
components with a scale height of about a hundred pc near the solar system
and a large filling factor.
Little is known about the distribution of the helium component because it
is only observable in one of its ionizing states.
It is assumed that its distribution closely follows that of hydrogen.

The correlation between the high-energy \gray{} intensity and the column
density of interstellar gas was well-established using the first \gray{} sky
surveys by SAS-2 and COS-B satellites \citep{LebrunPaul:1979,LebrunEtAl:1983},
and later confirmed by the EGRET telescope \citep{Hunter1997}.
The intrinsic connection between CRs and energetic \gray{s} inspired the
development of the first self-consistent model for CR propagation and diffuse
\gray{} emission and led to the establishment of the open-source
\GP\footnote{Available from
  \href{http://galprop.stanford.edu}{http://galprop.stanford.edu}\label{fn:link}}
  project in the mid-1990s
  \citep{MoskalenkoStrong:1998,StrongMoskalenko:1998,2000ApJ...528..357M,MoskalenkoStrong:2000,StrongEtAl:2000,VladimirovEtAl:2011}.
  Solving a system of $\sim90$ coupled transport equations and calculating the
  resulting high-energy emissions within a single framework enables the
  self-consistent treatment of all CR-related data.
  From the very beginning of
  the project \GP{} has been capable of full 3D spatial and time-dependent
  propagation calculations, but the data quality and computational requirements
  limited usage of these capabilities.
  Thus the CR propagation and production
  of secondary particles due to the interactions in the ISM relied on 2D
  cylindrical symmetric models for the respective spatial densities.

The data from the \fermilat{}, with its major improvements in sensitivity and
statistics compared to earlier experiments, heralded a new era for studies of
the high-energy interstellar emissions from the Milky Way.
\citet{AckermannEtAl:2012} considered a grid of 128 CR propagation models and
compared them against $\sim2$ years of \fermilat{} data to test the effects of
variations of important model parameters, such as the radial distribution of
the CR sources, the size of the propagation volume (the halo), and the spin
temperature of the \hi\ gas.
The models were all constructed using a 2D
(cylindrical symmetry) approximation for the CR propagation.
While the models provide reasonable agreement with the data, residuals
of the order of few tens of percent are visible on scales ranging from a few
to tens of degrees over the sky.
Some of these residuals are likely related to large-scale
structure in the CR and ISM distributions that is not described by
the 2D models, such as higher gas densities near the spiral arms and/or the presence of freshly accelerated CRs in the vicinity of their sources.
Consequently, analyses of the \fermilat{} data have inspired progress
towards more detailed 3D models for the high-energy interstellar
emissions: \citet{JohannessonEtAl:2013}, \citet{JohannessonEtAl:2015},
and \citet{PorterEtAl:2017} using \GP, \citet{KissmannEtAl:2017}
and \citet{NiederwangerEtAl:2017} with the {\it PICARD} code, and \citet{NavaEtAl:2017} with a Monte Carlo code.
Only for the \GP{}-based modeling has both the 3D structure of the ISM and CR spatial density distributions been taken into account; other works employ 2D ISM models.

Most of the knowledge of the gas distribution in the Milky Way has been acquired from line emission and absorption data.
For the \hi{} component the 21-cm hyperfine line is employed, which is
observed both in emission and absorption for a wide range of conditions
in the ISM \citep{DickeyLockman:1990, KalberlaKerp:2009}.
Using simplifying but realistic assumptions, the radiative transport equation for the line emission can be solved to directly relate the observations
of the emission line to the column density of the \hi{} gas \citep{KulkarniHeiles:1988}.
However, this requires information about the excitation temperature (hereafter, the so-called ``spin temperature'' $T_S$) of the emitting gas. 

The distribution of $T_S$ has been studied using observations of the 21-cm
line in absorption and found to range from a few tens of K to a few thousands
of K, and to be strongly correlated with the kinetic temperature of the
gas \citep{HeilesTroland:2003, StrasserTaylor:2004, DickeyEtAl:2009}.
This agrees with the idea that the \hi{} gas exists as two separate and stable phases in the ISM: the warm neutral medium (WNM, $T$$\sim$\,few thousand K) and the cold neutral medium (CNM, $T$$\sim$\,few tens of K).
The WNM has a larger filling factor and is generally more widely spread than the CNM that is more clumpy and has a smaller scale height, at least in the inner Galaxy.
Recent 21-cm absorption studies toward the outer Galaxy indicate that the warm and cold components are well mixed in that region, most likely because of the smaller amounts of molecular gas \citep{DickeyEtAl:2009}.

The distribution of H$_2$ is less well known because typical conditions in the
cold ISM do not produce detectable line emissions.
Other tracers must
therefore be used to estimate its column density.
The most common is the rotational transition line from the $^{12}$C$^{16}$O (hereafter CO) molecule, which is the second most abundant molecule in the ISM after H$_2$.
The formation conditions of CO are similar to those of H$_2$, and the line emissions are mostly excited through collisions between CO and H$_2$ molecules.
It has been observationally shown that the integrated line intensity of the CO lines is almost linearly related to the column density of H$_2$.
The linear conversion factor, $X_{\text{CO}}$ has been found to depend somewhat on both column density and temperature of the ISM \citep{BolattoEtAl:2013}.
Other molecular tracers, such as OH or $^{13}$CO in dense clouds, can also be used, but they are generally less abundant and their observations are more difficult. 

The Doppler shift of the line emission caused by differential movement of the interstellar gas in the Galaxy can be modeled to extract distance information.
The most common method assumes that the gas is in cylindrical rotation around the GC, which is a technique that has been applied since the beginning of systematic line-emission surveys \citep{Burton:1988, KalberlaKerp:2009}.
Even though this approach incorporates the main features of the gas motions, \citet{Burton:1988} pointed out that non-cylindrical streaming motions can cause significant perturbations to modeled line emission profiles.
These streaming motions have been shown to be up to 30~km~s$^{-1}$ using numerical simulations \citep{CheminEtAl:2015} and comparisons with other distance estimators \citep{TchernyshyovPeek:2017}.
Streaming motions dominate the line profiles in the directions toward the GC and anti-center where cylindrical rotation causes negligible motion along the line of sight (LOS).
In addition, thermal and turbulent motions cause line broadening at the level of a few km~s$^{-1}$, up to more than 10~km~s$^{-1}$ \citep{StrasserTaylor:2004}. 

The line broadening seriously affects the distance resolution available using the Doppler shift velocity, effectively smearing the gas along the LOS.
The resulting elongated features visible in many derivations of the Galactic distribution of interstellar gas \citep[e.g.,][]{NakanishiSofue:2003} are sometimes referred to as ``fingers of God'' because they all point towards the location of the Sun.
Some efforts have been made to correct for these inadequacies: \citet{LevineEtAl:2006} added elliptical rotation in the outer Galaxy to account for inconsistencies observed around the Galactic anti-center and \citet{KalberlaEtAl:2007} included decreased rotation for gas that is further above/below the Galactic disk, while \citet{PohlEtAl:2008} used hydrodynamical simulations to estimate the gas velocity fields to obtain distance estimates in the direction of the GC and used Gaussian profile fitting to account for the line broadening.
Even for the latter work noticeable artifacts are evident from the deconvolution procedure that smears out features in the actual spatial distribution.

Studies indicate that dust and gas in the ISM are well mixed
\citep[e.g.,][]{1978ApJ...224..132B}, and under certain assumptions, the dust
column density can also be used as a tracer of the gas column density
\citep[e.g.,][]{SchlegelEtAl:1998}.
More sensitive surveys of the stars in the
Galaxy over large areas of the sky can allow the 3D structure of the gas to be
probed through the observation of their light absorption by the dust, a
so-called dust reddening effect \citep[e.g.,][]{SchlaflyEtAl:2014}.
This
method can be more reliable than the kinematic distance estimators for the
emission lines and also works for directions toward the GC and anti-center.
However, its application is currently limited because it depends upon
observations of a large number of stars, and the light of more distant stars
is absorbed by the total column density along the LOS and, therefore, is very
faint.
Estimates of the gas distribution with this method require an
assumption about a conversion factor between the dust and gas column
densities that has been shown to be dependent on the physical properties of the ISM \citep[e.g.,][]{AdeEtAl:2015}. 

The techniques described above for deriving the distribution of the
interstellar gas in the Galaxy have been extensively applied
\citep[e.g.,][]{LevineEtAl:2006,PohlEtAl:2008,SchlaflyEtAl:2014,NakanishiSofue:2016,MarascoEtAl:2017,SchlaflyEtAl:2017}.
However, their results are not very well suited for usage as the gas distribution to run with CR propagation codes.
The major issue is that the distributions are usually incomplete with gaps along sight lines toward the GC; this is particularly prevalent for models derived via deconvolution of the line-emission survey data.
The stellar absorption method is limited by the distance from the Sun that can be probed (typically extending only out to $\sim10$ kpc), and by the sky coverage.
Even with the full-sky coverage of the \citet{PohlEtAl:2008} work, there are still issues with artifacts caused by broadening of the line emissions.

In this paper a forward
folding model fitting technique is employed to estimate the 3D structure
of the gas.
Continuity is enforced over the directions with limited distance
information by using a parameterized model of the gas distribution.
This approach resolves the issue of artifacts caused by smearing and allows for
complex gas rotation models that can also be parametrized and tuned to the
data.
It also allows the complexity of the spatial structure to be easily
controlled, as well as the
effects that each individual modification has on propagation of CRs to be studied separately.
The goal of this paper is to determine the distributions of \hi{}
and H$_2$ in the Milky Way, the most important gas components for modeling CR propagation,
production of secondaries, and high-energy interstellar emissions.
The \hii{}
gas, which has lower number density and larger scale height, is significantly
less important when modeling CR propagation and is not considered in this
paper.
The effects of 3D structure for the ISRF in combination with CRs and non-thermal interstellar emissions have been explored by \citet{PorterEtAl:2017}.

\section{3D modeling of the interstellar gas}
\label{sec:gasModeling}

\subsection{Analysis method}

A forward folding technique is used to derive
the 3D structure of the atomic and molecular hydrogen.
Parameterized models for both the gas densities and the velocity fields are tuned with a maximum-likelihood fit to \hi{} and CO line-emission data (described below).
The evaluations of the likelihood function are made with the \GG{} code, which is described in detail in Appendix~\ref{app:GALGAS}.
A brief overview of the code is given here.

The \GG{} code is designed to take arbitrary 3D models for the gas density and its velocity field and integrate them along the LOS from a user-specified location (the Solar system in this paper) to create line-emission profiles that can be compared to data.
The coordinate system is right handed with the Sun positioned at
the positive $X$-axis at a distance $R_\odot=8.5$~kpc from the Galactic center, while the $Z$-axis is pointing towards the north
Galactic pole. The $Z=0$ plane coincides with Galactic latitude $b=0^\circ$
and the $Y$-axis is parallel to the Galactic longitude $l=-90^\circ$.
In this coordinate system the conversion from $(l,b,s)$ to $(X,Y,Z)$ is given
by
\begin{align*}
  X &= R_\odot - s \cos(b)\cos(l),\\
  Y &= - s \cos(b)\sin(l),\quad \text{and}\\
  Z &= s \sin(b).
\end{align*}
where $s$ is the distance along the LOS $(l,b)$. 
For each LOS, the code uses the velocity field projected onto the line of
  sight to create a projection from  $s$ to Doppler-shifted
velocity $v$ which is calculated as the difference in velocity at the origin and a point along the LOS.
Using this projection from $s$ to $v$ the code integrates the gas density
  along the line of sight to calculate the column density of gas associated
with each velocity bin and converts it to line emission as described below.
The code accounts for turbulent and thermal motions by smoothing the resulting emission profiles with a Gaussian kernel.

Conversion of gas column densities for each velocity bin $v$ to an observed
line intensity is necessary for comparison with data.
For \hi{}, it is assumed that within each distance bin $s$ the gas is homogeneous with a
spin temperature $\bar{T}_S$ that is specified with a parametrized
distribution:
\begin{equation}
  \bar{T}_S(v) = \frac{\int_{v} T_S(X,Y,Z) ds}{\int_{v} ds}
\end{equation}
where the integration is performed over distance bin $s$ associated with
velocity bin $v$.  
The column density is turned into observed brightness using a
formula by \citet{KulkarniHeiles:1988}:
\begin{equation}
  T_b(v) = \left[ \bar{T}_S(v) - T_{0}(v) \right] \left[ 1 - e^{-\tau(v)} \right],
 \label{eq:HItransport}
\end{equation} 
where the optical depth is given by $\tau(v) = N_{\text{\hi}}(v)/C \bar{T}_S(v)$,
$N_{\text{\hi}}(v) = \int_v n_{\text{\hi}}(X,Y,Z) ds$ 
is the column density of hydrogen, and $C =
1.83\times10^{18}$~cm$^{-2}$~K$^{-1}$~(km~s$^{-1}$)$^{-1}$ is a constant.
If multiple distance bins align within the same velocity bin along a LOS, the optical depth
from distance bins between the observer and the current bin is also
included.
In this work the only background considered is the cosmic microwave
  background and the background temperature $T_{0}=2.66$~K is constant over
the sky.
The Galactic synchrotron continuum emission is non-negligible at
1420~MHz. Its 3D distribution is not well known and is difficult to
explicitly account for in Eq.~(\ref{eq:HItransport}), and is therefore not included.
Typical estimates \citep{Sofue:2017} will lead to the emission from \hi{} from
the Galactic disk being underestimated by $\sim 10$~K for the model
presented in this paper. Meanwhile, the optically thin assumption is
used when modeling the \hi{} gas, which is implemented using a large
constant value for $T_S(X,Y,Z)$. This assumption has a much larger effect on
the estimated emission than neglecting the synchrotron continuum
emission. Hence, the analysis in this paper provides a robust lower
limit for the density distribution of \hi{} gas.
  
For the molecular gas, the standard assumption that
the column density of H$_2$ is linearly related to the integrated line
emission of CO is used, $N_{\text{H}_2}(v) = X_{\text{CO}}(v)
W_{\text{CO}}(v)$. Here $W_{\text{CO}}(v)$ is the integrated CO line emission
over the velocity bin $v$ and $X_{\text{CO}}(v)$ is the linear conversion
factor.
This assumption of linear relation is entirely phenomenological and does not
have a strong physical motivation.
The CO emission is generally optically thick and the linear relation between
$N_{\text{H}_2}$ and $W_{\text{CO}}$ may be caused by the prevalent
conditions in the ISM \citep{GloverMacLow:2011}.

This assumption leads to a simple relation for the number density of H$_2$:
\begin{equation}
 n_{\text{H}_2}(X,Y,Z) = X_{\text{CO}}(X,Y,Z) \epsilon_{\text{CO}}(X,Y,Z)
 \label{eq:H_2COrelation}
\end{equation}
where the linear conversion factor $X_{\text{CO}}(X,Y,Z)$ can depend
on the position in the Galaxy and $\epsilon_{\text{CO}}$ is the CO volume
emissivity.
This greatly simplifies the modeling of the CO line emission because the
quantity of interest becomes $\epsilon_{\text{CO}}(X,Y,Z)$, which is
independent of $X_{\text{CO}}(X,Y,Z)$ and linearly related to
the CO line emission,
\begin{equation}
  W_{\text{CO}}(v) = \int_v \epsilon_{\text{CO}}(X,Y,Z) ds
\end{equation}
However, propagation codes require the number density of the hydrogen gas
and specification of $X_{\text{CO}}(X,Y,Z)$ is necessary.
Common assumptions include a constant $X_{\text{CO}}$ throughout the Galaxy
or a radially increasing $X_{\text{CO}}(R)$ that seems to be more consistent
with \gray{} data \citep{AckermannEtAl:2012}.

The model parameters are tuned by maximizing the likelihood of the model given the data.
Even though the data is assumed to be normally distributed with a specified uncertainty, a student-t likelihood is used.
The student-t distribution has more weight in the tails of the distribution compared to the normal distribution and the likelihood is less affected by strong outliers in the data.
This property of the likelihood is desired because the models used in the analysis are by design not able to recover the fine structure of the gas
distribution.
Generally, there are more data points with low or little emission than bright
ones, so using the student-t likelihood de-weights the emission peaks allowing a simplified model to catch the basic features of the data without being biased by strong emission peaks that it cannot reproduce.
Also, the emission peaks not accounted for by the model will be evident in the residuals for the longitude and latitude profiles, making it easier to identify were additional model refinement is needed.
Using the student-t likelihood requires, in addition to the data uncertainty, specifying the number of degrees of freedom which is set to $\nu=100$ for this analysis.
The exact value does not change the overall conclusions of this work, but smaller values generally result in smaller estimated total gas mass with comparatively smaller negative residuals, while larger values give higher gas masses with larger negative residuals.
Comparison with likelihoods using normally distributed errors show that $\nu\gtrsim 10^4$ is needed before the student-t likelihood gives similar results.

\subsection{Data}

This paper employs the \hi{} LAB survey \citep{KalberlaEtAl:2005} and the
composite CO survey of \citet{DameEtAl:2001} for the model tuning. The data is
re-binned to a HEALPix grid \citep{GorskiEtAl:2005} using HEALPix order 7 for
\hi{} while order 8 is used for the CO data. The selected resolution on the
sky is such that the spatial resolution at the GC is about 80~pc and 40~pc,
respectively for CO and \hi{}. This is enough to resolve the gradient of the
gas distributions near the GC.
The velocity resolution of both surveys is degraded to 2~km~s$^{-1}$ velocity bins to reduce the needed computational resources.
The lower velocity resolution does not strongly affect the results because it is
still smaller than the characteristic line spread for CO and \hi{}, which
is assumed to be 6~km~s$^{-1}$ and 10~km~s$^{-1}$, respectively.
These values
were determined by examining the tail of the line emission close to the
tangent point velocities in the inner Galaxy and are in reasonable agreement
with the analysis of \citet{MarascoEtAl:2017}.
Non-cylindrical motions of the
gas dominate the Doppler shifts of the line emission in directions towards the
GC and anti-center so the velocity information is ignored for longitudes $|l|
< 10^\circ$ and $170^\circ < l < 190^\circ$ and only the total velocity
integrated emission is compared.
In those regions the uncertainty for the bins along each LOS is summed up in quadrature assuming
that the uncertainties are independent.

The statistical uncertainty on the data is assumed to be constant over the
entire sky.
Values of 0.05~K for CO and 0.1~K for \hi{} are used, which are consistent with the noise estimate for the original
surveys taking into account the re-binning.
Because the models are too simple
to properly account for the fine structure in the data the statistical
uncertainty of the model parameters is not of high importance and therefore neither is the exact value of the uncertainty on the data.
  However, some of the model
parameters described below are common in both the CO and \hi{} models and
constrained by both data sets and it is thus important to have the
relative uncertainty of the two datasets correct to avoid biasing the
likelihood either way.
The statistics of the two data sets are such that the much brighter \hi{}
emission dominates the likelihood.

The \hi{} data is filtered to exclude high-velocity emission, and also emission from the Local Group galaxies (Large and Small Magellanic Clouds, M31, and
M33).
No attempt is made to correct for bright radio background sources because the resolution of the LAB survey is not sufficient to do that accurately.
Only data with $|b|\leq 40^\circ$ is selected for the analysis because higher latitude emission is predominantly due to local clouds that are not elements of the models considered in this paper.
The increased number of pixels used for the likelihood evaluation from including high-latitude data considerably increases the required computation without providing additional constraints for the parameters of the global model.
The CO survey is filtered with the moment masking method of \citet{Dame:2011}, significantly reducing noise in the data.
No other filtering is performed on the CO data.

\subsection{Model Components}

The model components used in this work are: a warped disk with a scale height
that varies with Galactocentric radius, a central bulge/bar, and 4 logarithmic
spiral arms.
To reduce the number of parameters many of the geometrical
parameters are the same for both the \hi{} and CO models.
This includes the
parameters controlling the warp of the disk, the radial increase in scale
height, the shape of the bulge/bar, and the shape of the spiral arms.
The
physical motivation for this assumption is that these parameters are
controlled by external processes and should affect both components of the gas
in a similar way.
The radial and vertical profiles of the spiral arms follow
that of the disk to further reduce the number of parameters.
The velocity
field is modeled as cylindrical rotation using the rotation curve of
\citet{SofueEtAl:2009} scaled to match the IAU-recommended Sun-GC distance of
$R_\odot=8.5$~kpc and the rotation velocity of $v_\odot = 220$~km~s$^{-1}$ at the location of the
solar system.
The
distance from the GC projected onto the $X$-$Y$ plane is calculated as $R = \sqrt{X^2+Y^2}$.
The optically thin assumption is used for the \hi{} data, which is effectively modeled using a large constant value for $T_S$ in Eq.~(\ref{eq:HItransport}).
This provides a robust lower limit for the column density of \hi{} gas.

Two functional forms of the radial profile for the number density of the disk and spiral arms are explored, an exponential disk with a central hole
\begin{equation}
 f_d (R) = n_d e^{-(R-r_s)/r_0} \left[ 1 - e^{\left( -R/r_h \right)^{h_i}} \right], 
 \label{eq:ExpDisc}
\end{equation}
where $r_s=8.0$ kpc\footnote{$r_s$ is a normalization constant and its value
  was chosen to coincide with the point in the cubic spline closest to
$R_\odot$}, 
$n_d, r_0, r_h, h_i$ are free parameters, and a cubic
spline in logarithm of the number density $n(R)$ between the values at
constant radii, $R= 0, 2, 4, 6, 8, 10, 15, 20$, and $50$ kpc.
The radii are selected based on previously determined radial profiles for the \hi{} and CO gas while minimizing the number of parameters \citep{GordonBurton:1976,BronfmanEtAl:1988}.
The cubic spline provides more freedom but at the expense of more than double the number of parameters.
When combined with the bulge/bar component described below, the number density at the two innermost radial points of the cubic spline is fixed to a small value.

The warp of the Galactic disk is modeled similarly to \citet{LevineEtAl:2006}:
\begin{equation}
 z_0 = w_0(R) + w_1(R)\sin\left( \Theta - \theta_1 \right) +
 w_2(R)\sin\left( 2\Theta - \theta_2 \right)
 \label{eq:warp}
\end{equation}
where $\Theta = \tan^{-1}(Y/X)$ is the azimuthal angle.
The radial dependence
of the amplitudes $w_i(R)$ is modeled with a cubic spline between the constant
radii: 0, 5, 10, 15, 20, and 50 kpc.
While the warp is dominantly in the outer
Galaxy the points in the inner Galaxy are used to account for small variations
in the disk mid-plane as found by \citet{BronfmanEtAl:1988}.
The zero modes,
$\theta_i$, are assumed to be independent of the radius.
The vertical profile
of the disk is modeled as a function of $Z'/z_h$, where $Z' = Z - z_0$ is the
distance from the central plane of the disk, and $z_h$ is the scale height of
the disk.
Several functions describing the vertical
profile are tested, including an exponential, a Gaussian, and hyperbolic
secant to the power of $2, 1$, and $0.5$.
To account for the increasing scale
height in the outer Galaxy, its radial dependence is modeled as 
\begin{equation}
 z_h(R) = z_s e^{(R-r_{z_0})/r_z},
 \label{eq:scaleHeight}
\end{equation}
where $r_{z_0}=8.5$ kpc is a constant\footnote{$z_s$ is thus the scale height of the gas at
the solar location.}, and $z_s, r_z$ are free parameters.
The final disk model is thus
\begin{equation}
 f_d(X,Y,Z) = f_d(R) f_s(R,Z'),
 \label{eq:totalDisk}
\end{equation}
where $f_s(R,Z')$ is one of the scale functions describing the vertical profile of the disk.

The central bulge/bar component is parameterized with the function
\begin{equation}
 \begin{aligned}
 f_{b}(X,Y,Z) &= n_{b} e^{-R_r^{e_i}} R_r^{p_i} \quad \text{with}\\
 R_r &= \left( R'/r_b + Z/z_b \right)^{-1}, \\
 R' &= \sqrt{ \left( X' \right)^2 + \left( Y'/0.3 \right)^2}, \\
 X' &= X \cos(\theta_b) + Y \sin(\theta_b) + x_0, \\ 
 Y' &= -X \sin(\theta_b) + Y \cos(\theta_b).
 \label{eq:Bulge}
 \end{aligned}
\end{equation}
where $n_{b}$, $e_i$, $p_i$, $r_b$, $z_b$, and $x_0$ are free parameters while
$\theta_b = -30^\circ$ is a constant.
The lack of velocity and, therefore, distance
information means that it is not possible to constrain $\theta_b$ using the
method employed in this paper.
The exact value chosen for $\theta_b$ will
affect the other parameters of the bulge, but not the data-model agreement.
The value of $\theta_b$ that we use is within the range of $-10^\circ$
\citep{Freudenreich:1998} to $-45^\circ$ \citep{Lopez-CorredoiraEtAl:2007} given in the literature.
The rotation of the bulge/bar corresponds to its closest distance being at positive longitudes. 

\begin{deluxetable}{lc}[tb!]
 \tablecaption{\label{tab:diskParameters} Model parameters describing the
 radial and vertical distributions of the number density for the \hi\ and CO
 models}
 \tablecolumns{2}
 \tablehead{\colhead{Parameter} & \colhead{Value} }
 \startdata
 \multicolumn{2}{l}{\sc \quad Disk parameters for CO model} \smallskip\\
 $n_{d}$,  K~km~s$^{-1}$~kpc$^{-1}$ & 0.894 \\
 $r_s$\tablenotemark{a}, kpc & 8.0 \\
 $r_0$, kpc & 1.27 \\
 $r_h$, kpc & 6.34 \\
 $h_i$, kpc & 6.38 \\
 $z_{s,\text{CO}}$\tablenotemark{b}, kpc & 0.103 \medskip \\
 
 \multicolumn{2}{l}{\sc \quad Disk parameters for \hi\ model} \smallskip\\
 $n(R=8\,\text{kpc}) \equiv n_8$, cm$^{-3}$ & 0.160 \\
 $n(0\,\text{kpc})/n_8$ & $1.72 \times 10^{-6}$ \\
 $n(2\,\text{kpc})/n_8$ & $0.284 $ \\
 $n(4\,\text{kpc})/n_8$ & $0.807$ \\
 $n(6\,\text{kpc})/n_8$ & $1.19$ \\
 $n(10\,\text{kpc})/n_8$ & $0.798$ \\
 $n(15\,\text{kpc})/n_8$ & $0.477$ \\
 $n(20\,\text{kpc})/n_8$ & $0.0457$ \\
 $n(50\,\text{kpc})/n_8$ & $1.12 \times 10^{-3}$ \\
 $z_{s,\text{\hi}}$\tablenotemark{b}, kpc & 0.0942 \medskip \\
 
 \multicolumn{2}{l}{\sc \quad Bulge parameters for CO model} \smallskip\\
 $n_{b}$, K km s$^{-1}$ kpc$^{-1}$ & 47.8 \\
 $\theta_b$\tablenotemark{a}, rad & 5.67 \\
 $x_0$, kpc & 0.751 \\ 
 $r_b$, kpc & 0.514 \\
 $z_b$, pc & 6.43 \\
 $e_i$ & 0.647 \\
 $p_i$ & 1.18 \medskip \\
 
 \multicolumn{2}{l}{\sc \quad Flare parameters (common)} \smallskip\\
 $r_z$, kpc & 6.94 \\
 $r_{z_0}$\tablenotemark{a}, kpc & 8.5 
 \enddata
 \tablenotetext{a}{Constant}
 \tablenotetext{b}{Note that the CO model uses squared hyperbolic secant,
 while the \hi\ model uses the square root of hyperbolic secant for the
 vertical scale. These numbers are therefore not directly comparable.}
\end{deluxetable}

\begin{deluxetable}{lccc}[tb!]
 \tablecaption{\label{tab:warpParameters}Model parameters describing the disk warp}
 \tablecolumns{4}
 \tablehead{\colhead{Parameter} & \colhead{Mode 0} & \colhead{Mode 1} &
 \colhead{Mode 2} }
 \startdata
 $\theta$, rad & \nodata & 4.61 & 2.73 \\
 $w(0\,\text{kpc})$, kpc & --0.0756 & 0.146 & 0.287 \\
 $w(5\,\text{kpc})$, kpc & --0.00819 & --0.0520 & --0.0192 \\
 $w(10\,\text{kpc})$, kpc & --0.0288 & 0.101 & --0.00716 \\
 $w(15\,\text{kpc})$, kpc & 0.0576 & 0.737 & --0.00587 \\
 $w(20\,\text{kpc})$, kpc & 0.767 & 1.71 & 0.587 \\
 $w(50\,\text{kpc})$, kpc & 20\tablenotemark{a} & 20\tablenotemark{a} & 14.9 
 \enddata
 \tablenotetext{a}{Parameter at fit range boundary.}
\end{deluxetable}

\begin{deluxetable}{cccccc}[b!]
 \tablecaption{\label{tab:armParameters}Model parameters describing the shape and number density of the spiral arms}
 \tablecolumns{6}
 \tablehead{
 Arm & \colhead{$\alpha_j$} & \colhead{$r_{\text{min},j}$} &
 \colhead{$\theta_{\text{min},j}$\tablenotemark{a}} &
 \colhead{$\epsilon_{\text{CO}}(8\,\text{kpc})$} & \colhead{$n_{\text{\hi}}(8\,\text{kpc})$} \\
 No. & & kpc & rad & K\,km\,s$^{-1}$\,kpc$^{-1}$ & cm$^{-3}$
 }
 \startdata
  1 & 3.30 & 2.00 & 1.05 & 0.642 & 0.184 \\
  2 & 4.35 & 3.31 & 2.62 & 0\tablenotemark{b} & 0.193 \\
  3 & 5.32 & 3.89 & 4.19 & 3.37 & 0.332 \\
  4 & 4.75 & 3.19 & 5.76 & 7.53 & 0.521 
 \enddata
 \tablenotetext{a}{Constant}
 \tablenotetext{b}{Parameter at fit range boundary.}
\end{deluxetable}

The spiral arms are purely logarithmic: 
\begin{equation}
 \theta_j(R) = \alpha_j \log\left( R/r_{\text{min},j} \right) +
 \theta_{\text{min},j},
 \label{eq:arms}
\end{equation}
where $\alpha_j$, $r_{\text{min},j}$ and $\theta_{\text{min},j}$ are
parameters of the model.
Values of $\alpha_j$ and $r_{\text{min},j}$ are tuned
in the minimization procedure, while the starting angles
$\theta_{\text{min},j} = -\pi/6 + j\pi/2$ are held constant throughout.
The pitch angle of the spiral arm can be determined using $\theta_{p,j} = \tan^{-1}(\alpha_j^{-1})$.
The starting point of arms 2 and 4 are at the ends of the central bulge/bar.
The arms have a Gaussian profile perpendicular to the locus traced by Eq.~(\ref{eq:arms}) with a scale of 0.6~kpc giving them a FWHM of $\sim 1.4$~kpc.
The radial density distribution of each arm is assumed to be identical to that of the disk in each density model, but with independent normalization for each arm.
The vertical scale height of the arms is the same as that of the disk for the respective gas component.

To ensure stable fits, the maximum-likelihood procedure is performed in several iterations using the best-fit values from the previous step as a starting point for the next.
The initial fit is performed using the disk component only without the warp and radial increase in the scale height.
The model complexity is then increased by fitting for additional parameters and components in the following order: the radial distribution of the vertical scale height, the central bulge/bar, the warp of the disk, and the spiral arms.
Because of the number of parameters and computing time required, the selection for the different radial and vertical profiles is performed without inclusion of the warp and spiral arms.

\subsection{Results}

According to the maximum-likelihood values, the logarithmic cubic-spline
radial profile is a better match for the \hi{} number density profile, while
the exponential disk with a central hole is preferred for the CO distribution.
The central bulge/bar component is rejected by the fit for the \hi{} model and subsequently excluded from the \hi{} model.
The best-fit vertical profile is the square root of the hyperbolic secant for the \hi{} model, while for CO it is the square of the hyperbolic secant that gives the best fit of all the tested profiles.

The parameters and their best-fit values for the final models of \hi{} and CO
gas are listed in Table~\ref{tab:diskParameters} for the radial and vertical
number density distributions of the disk, in Table~\ref{tab:warpParameters}
for the warp describing the central plane of the disk, and in
Table~\ref{tab:armParameters} for the spiral arm parameters.
Because the
combined model does not reproduce the fine structure of the data, statistical
uncertainties are unimportant and not reported.
The statistical
uncertainties are in most cases less than 0.1\%.
Each parameter value is
reported with 3 significant digits, or at the level of the statistical
uncertainty, whichever has a fewer number of significant digits.
The final
models will be distributed with larger number of significant digits as
supplementary material to the paper in XML form readable by the
\GP{}\textsuperscript{\ref{fn:link}} code version 56.

\begin{figure*}[tb!]
 \centering
 \includegraphics[width=.50\textwidth]{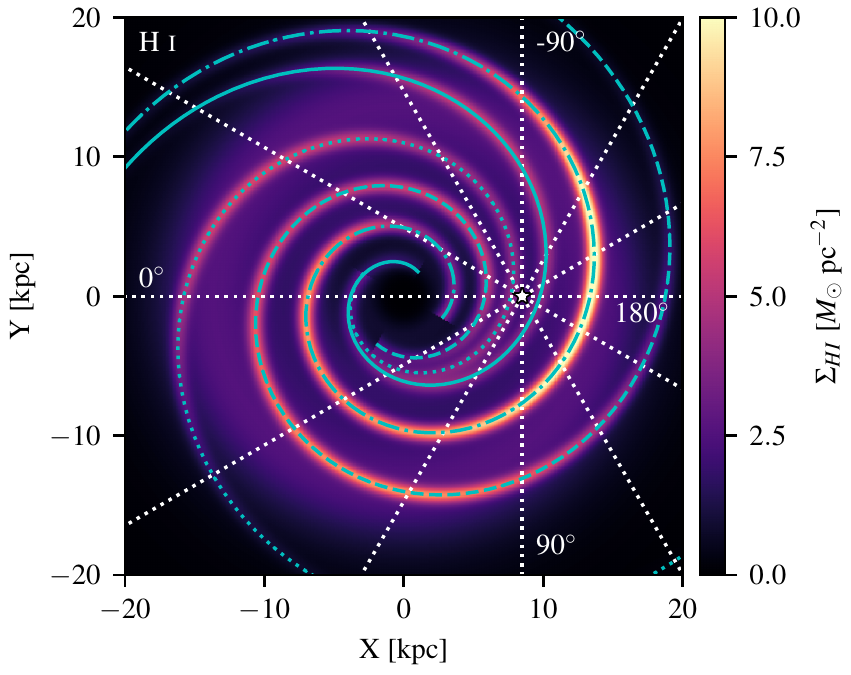}\hfill
 \includegraphics[width=.50\textwidth]{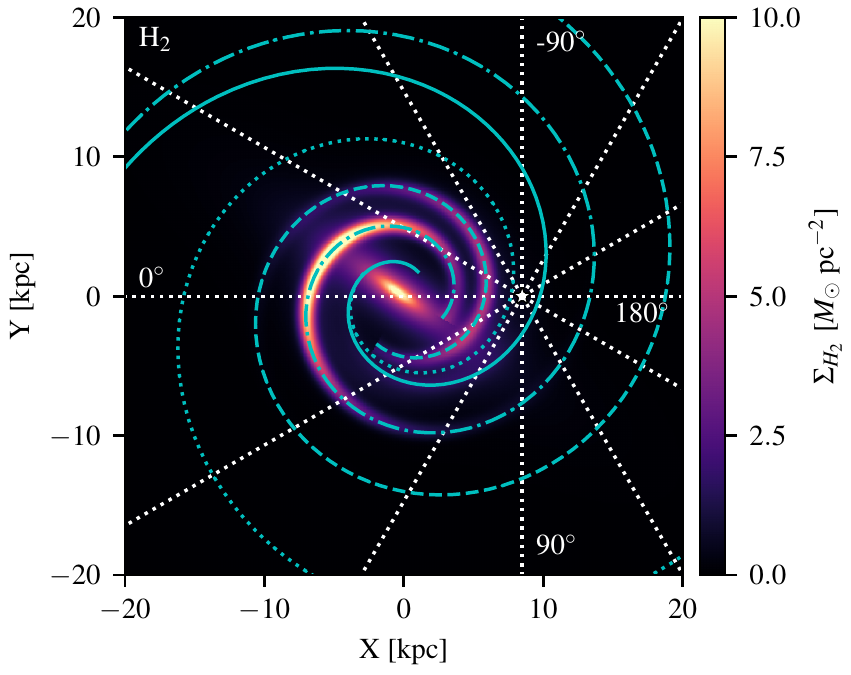}\\
 \includegraphics[width=.50\textwidth]{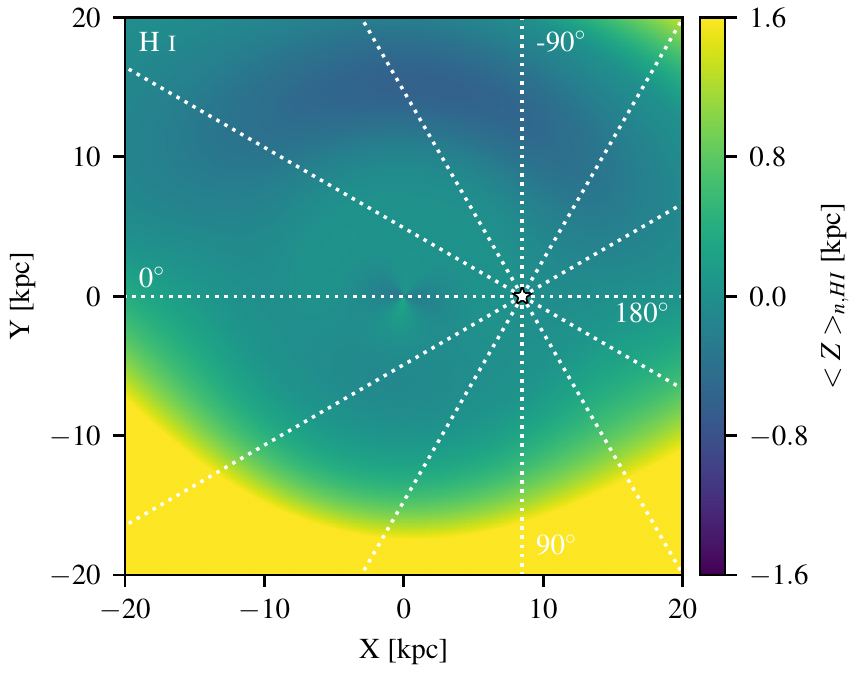}\hfill
 \includegraphics[width=.50\textwidth]{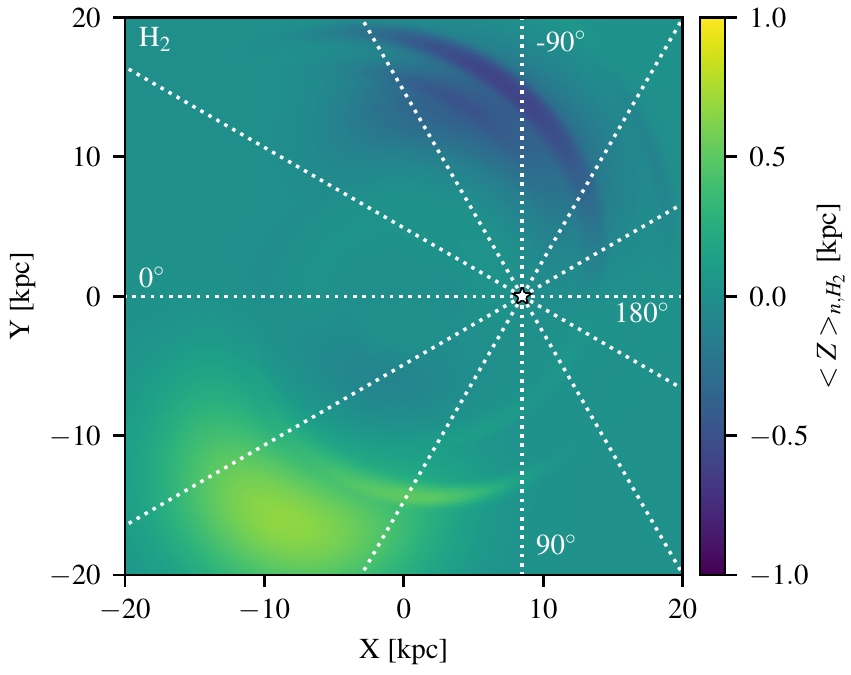}
 \caption{Surface density maps for the final models (top) and maps of the first moment of the vertical density distribution (bottom). \hi{} gas component is on left and H$_2$ component is on right. The CO number density is converted to H$_2$ number density assuming $X_{\rm CO} = 2 \times 10^{20}$ cm$^{-2}$ (K km s$^{-1}$)$^{-1}$. The Sun is marked as a white point and the white-dashed lines mark the longitude grid with 30$^\circ$ step. The cyan curves on the surface density maps trace the cores of the spiral arms with each arm marked with a different line style: arm 1 is solid, arm 2 -- dotted, arm 3 -- dashed, and arm 4 -- dash-dotted.}
 \label{fig:gasMaps}
\end{figure*}

The best-fit model parameters are in reasonable agreement with previous
studies of the interstellar gas.
The radial distribution of the \hi{} surface
density is approximately flat between 4 and 15~kpc \citep{GordonBurton:1976}.
The distribution then falls off exponentially towards the outer Galaxy at a rate
similar to that found by \citet{KalberlaDedes:2008}.
The CO distribution has a peak between 4 and 6~kpc \citep{BronfmanEtAl:1988} before falling off exponentially at larger Galactocentric distances.
The central bulge/bar accounts for some part of strong CO emission in the inner Galaxy \citep{FerriereEtAl:2007}.
The scale parameter for the disk flaring is somewhat smaller than that derived for the global average by \citet{KalberlaDedes:2008}.
The flaring parameter value is found to be closer to its value for the northern Galactic hemisphere indicating that northern latitudes ($b > 0^\circ$) may have more weight when fitting the parameters of the described model.
The warp parameters are in excellent agreement with those found by \citet{LevineEtAl:2006} for Galactocentric radii between 10 and 25~kpc.
The best-fit arm pitch angles are $16.8^\circ$ (arm 1), $12.9^\circ$ (arm 2), $10.6^\circ$ (arm 3), and $11.9^\circ$ (arm 4).
These are all within the range of arm pitch angle estimates from the literature \citep[e.g.,][]{Vallee:2017}.

The total masses of \hi{} and H$_2$ gas components in the constructed models
are $4.9 \times 10^9$~M$_\odot$ (\hi{}) and $0.67 \times 10^9$~M$_\odot$
(H$_2$), where the standard value of $X_{\rm CO}
= 2 \times 10^{20}$ cm$^{-2}$ (K km~s$^{-1}$)$^{-1}$ \citep{BolattoEtAl:2013}
is used for the conversion.
The spiral arms in the \hi{} model account for approximately 25\% of the mass
while the remainder is in the disk. For the CO model the arms account for
$\sim42$\% of the mass, the bulge/bar $\sim31$\%, and $\sim27$\% is in
the disk component.
These mass ratios may not relate directly to an estimated
H$_2$ mass because the $X_{\text{CO}}$ conversion factor has been shown to
depend on the specific properties of the ISM, and near the GC it may be an
order of magnitude smaller than the Galactic average \citep{FerriereEtAl:2007,AckermannEtAl:2012}.
For example, using $X_{\rm CO} = 2 \times 10^{19}$ cm$^{-2}$ (K km~s$^{-1}$)$^{-1}$ for the bulge/bar component
reduces the total H$_2$ mass in the model to $0.48 \times 10^9$~M$_\odot$ and
the fraction of the mass in the bulge/bar to $\sim5$\%. 

\begin{figure*}[tb!]
 \center{
 \includegraphics[width=0.5\textwidth]{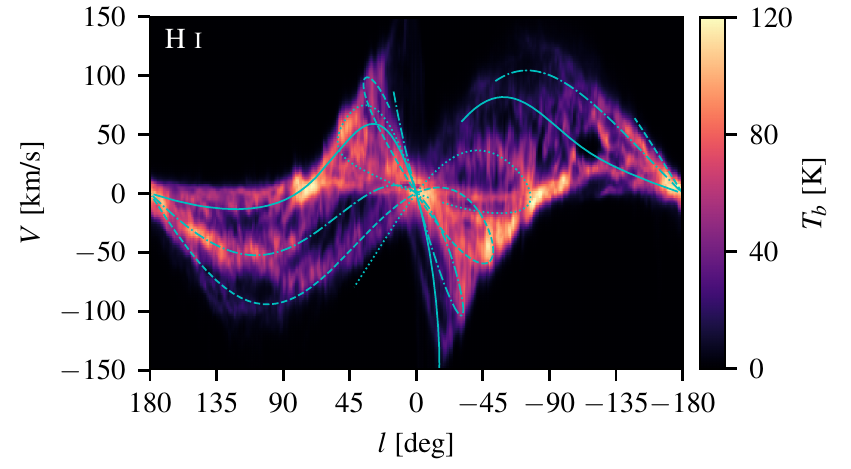}\hfill
 \includegraphics[width=0.5\textwidth]{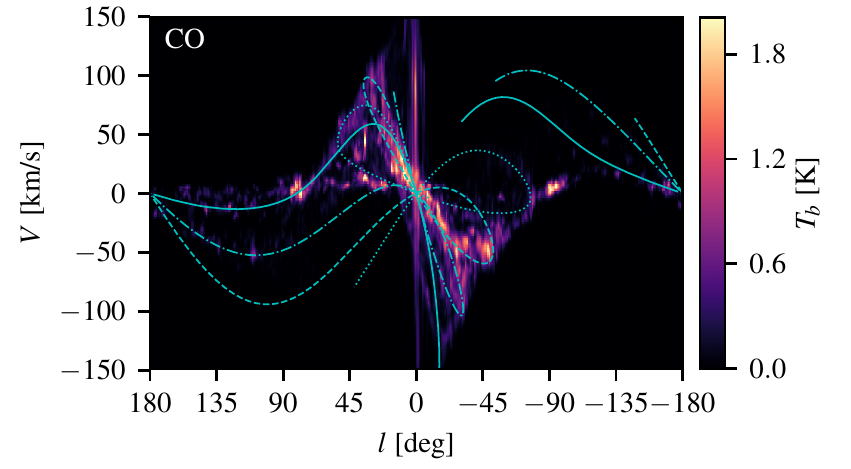}\\
 \includegraphics[width=0.5\textwidth]{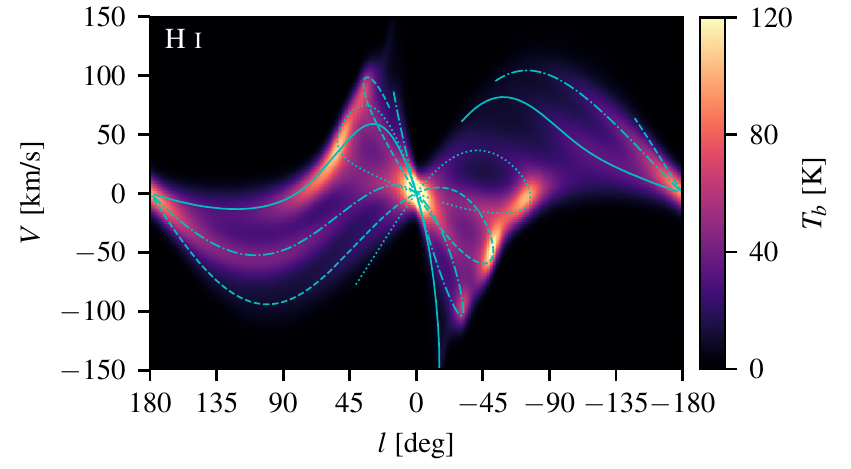}\hfill
 \includegraphics[width=0.5\textwidth]{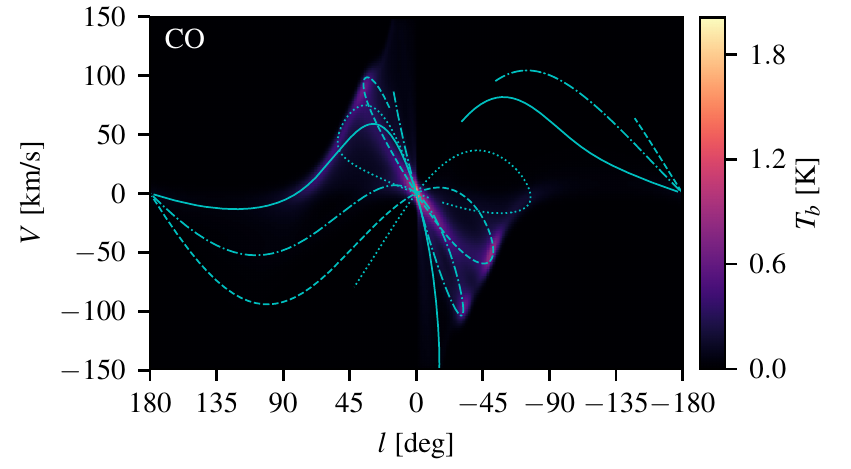}
 }
 \caption{Longitude-velocity diagram integrated over the longitude range $|b|
   < 4^\circ$ for CO (left panels) and \hi\ (right panels) gas. Top row shows
   the data, while bottom row shows our best-fit model. The cyan curves trace
   the cores of the spiral arms, the line coding is the same as in Figure~\ref{fig:gasMaps}.
 }
 \label{fig:lvdiagram}
\end{figure*}

The surface density maps of the final models
\begin{equation}
  \Sigma (X,Y) = \int_{-\infty}^\infty n(X,Y,Z) dZ
 \label{eq:surfaceDensity}
\end{equation}
are shown in Figure~\ref{fig:gasMaps}.
The CO model has been converted to H$_2$ surface density using $X_{\text{CO}} = 2 \times 10^{20}$ cm$^{-2}$ (K~km~s$^{-1}$)$^{-1}$.
The central bulge/bar and some spiral arms are clearly visible on the H$_2$ surface density map, meanwhile, the lack of arm 1 and the faint haze at the location of arm 2 are also apparent.
Comparison of the spiral arm structures to the catalog of the spiral arm tangents collected by \citet{Vallee:2014} leads to the following identifications:  arm 1 accounts
for the Perseus arm, arm 2 -- for Sagittarius and Carina, arm 3 -- for Scutum
and Crux-Centaurus, and arm 4 -- for the 3~kpc arm, Norma, and Cygnus.
Arm 3
also accounts for the new arm detected by \citet{DameThaddeus:2011}, which has
been confirmed using parallax observations by \citet{SannaEtAl:2017}.
This picture is also mostly consistent with the updated spiral arm model presented by \citet{Vallee:2016}.
The starting points of the arms shown in his Figure~2 are at 2~kpc so care must be taken when matching the arms. The only exception is that arm 1 does not match the Perseus arm in the updated model of \citet{Vallee:2016}. 

The final arm parameter values are somewhat dependent on their initial values
chosen for the model fit and their interpretation could be ambiguous.
In particular, arm 1 does not match perfectly with the start of Perseus at
$l=-23^\circ$; its start is closer to $l=-15^\circ$.
It also overlaps with arm 2 in the inner Galaxy, indicating that there is some
degeneracy between these arms.
In turn, this may affect the derived values of the $\alpha_j$ parameters so the modeled spiral arms match observations in the outer Galaxy.
The presented spiral arm configuration is quite stable over the relatively few choices of initial values tested in this analysis.
The resulting small variations in the model parameters do not qualitatively affect the results for CR propagation and high-energy interstellar emission
(Section~\ref{sec:CRgamma}) and further exploration of the model is
deferred to future work.  

To illustrate the warp of the Galactic disk, the first mode of the calculated vertical number density distribution
\begin{equation}
  \left<Z\right>_n(X,Y) = \frac{ \int_{-\infty}^{\infty} Z n(X,Y,Z) dZ}{\Sigma(X,Y)}
 \label{eq:meanHeight}
\end{equation}
is shown in Figure~\ref{fig:gasMaps}.
The results are in good agreement with those from \cite{LevineEtAl:2006}
(their Figure~11).
In this work the same warp is applied to the disk and spiral arm components of the \hi{} and CO models, but the central bulge/bar does not include any warp.
The best-fit
parameters for the CO model correspond to the bulge/bar component that falls
off with the Galactocentric distance slower than the disk and spiral arm
components.
The bulge/bar contributes to the gas
number density even outside the solar radius and the warp maps of the
\hi{} and CO models are not identical even though
the warp component is the same for both models.
The contribution of the bulge/bar component falls off more quickly along its minor axis in the plane resulting in larger values of the first mode along that direction.
The warp is not significant for the CO model because its number density for distances $\gtrsim 10$~kpc is low.
Even though the effect of the warp is small in the inner Galaxy it is found to be necessary for the model because it significantly improves the data-model agreement.

Figure~\ref{fig:lvdiagram} shows the longitude-velocity diagram for both models and data integrated over the Galactic disk $|b| < 4^\circ$, with the locations of the spiral arms overlaid.
Overall, the models reproduce the data fairly well, with the main ``butterfly'' shape of the model driven by the cylindrical rotation.
The spiral arm structures are clearly visible in the \hi{} model plot and their locations reasonably match similar structures in the data.
However, the enforced smoothness of the models means that the true complexity of the observed ISM is not fully recoverable and even some large-scale features are not reproduced.
There is a clear spur between the spiral arm structure visible to the right of $(l,V)=(100^\circ, -60$ km~s$^{-1})$ in the \hi{} data that is absent in the model.
Gaps in the data near $(50^\circ, -30$ km~s$^{-1})$ and $(-110^\circ, 30$ km~s$^{-1})$ show the absence of the gas, but correspond to the spiral arms in the model.
Very bright emission near $(80^\circ, 0$ km~s$^{-1})$ and $(-90^\circ, 0$ km~s$^{-1})$ is not reproduced by \hi{} nor CO models.
The CO model is also much fainter than the data toward the inner Galaxy, and the few clouds visible in the outer Galaxy are not reproduced either.

Figure~\ref{fig:lvdiagram} also illustrates why the densities in spiral arms 1 and 2 are lower than in the other arms, especially for the CO model.
Arm 1 (associated with the Perseus arm and shown as a solid curve) starts in a void in the data at around $l=15^\circ$.
It then aligns with the location of arm 2 in a region with data and follows to some extent other evident features all the way to the outer Galaxy.
It looks shifted relative to the brightest features in the data at $90^\circ < l < 150^\circ$ in both CO and \hi{}.
Arm 1 ends up in a large void in the \hi{} data at $l \sim -110^\circ$ before picking up some structure further on.
There is also a bright feature near $(40^\circ, 40$ km~s$^{-1})$ in the CO data with corresponding structure in the \hi{} data that may be associated with this spiral arm.
The location of this feature is, however, offset from the modeled spiral arm. Arm 2 that has been associated with Sagittarius and Carina and shown as a dotted curve seems to be offset from a very bright feature in the \hi{} data at $15^\circ < l < 45^\circ$, and from a fainter feature at $-70^\circ < l < -20^\circ$.
There is evidence of similar offsets in the CO data as well.
The offsets mentioned above are likely caused by some combination of an incorrect spiral arm shape and/or variations in the velocity field.
However, it is difficult to discriminate between these two effects without additional information.

\begin{figure}[tb!]
 \centering
 \includegraphics[width=0.48\textwidth]{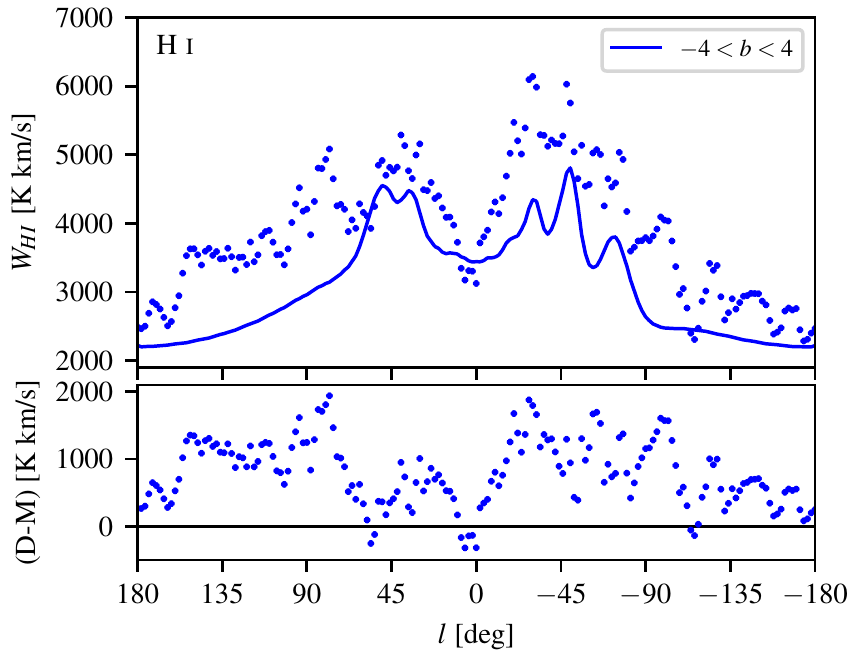}\\
 \includegraphics[width=0.48\textwidth]{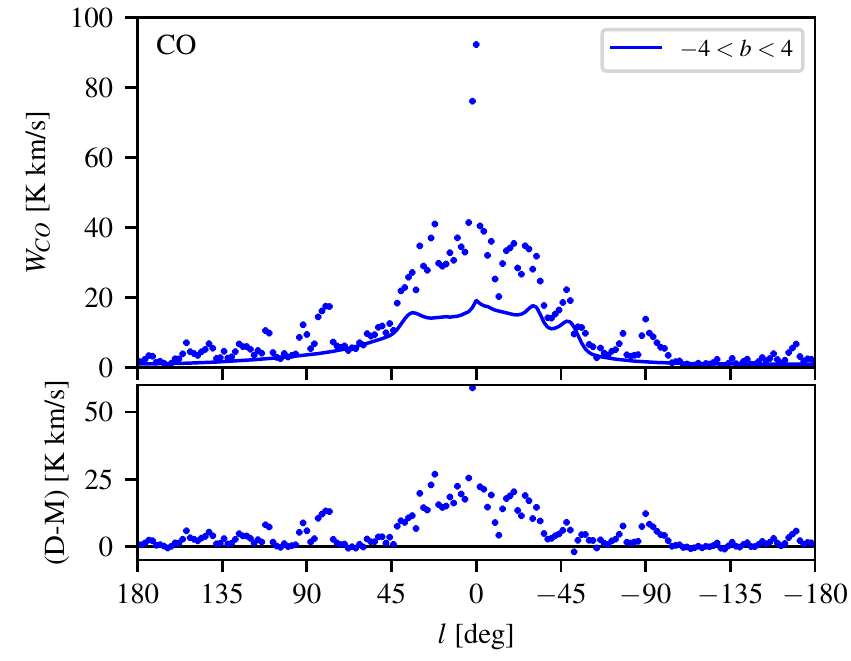}
 \caption{Longitude profiles of the models (curves) overlaid on the data
 (points). The model and data are integrated over all velocity bins and
 averaged over the latitude range $|b| < 4^\circ$. \hi{} gas is on top and CO is at the bottom.
 }
 \label{fig:longitudeProfiles}
\end{figure}

The longitude profiles (Figure~\ref{fig:longitudeProfiles}) show that the data
are under-predicted by the models.
This is a consequence of using the
student-t likelihood, which de-weights strong outliers in the data, in
combination with a simplified and smooth model.
The outliers are positive in
almost all cases giving rise to positive residuals.
Of the two, the \hi{} model performs somewhat better at representing the data with the
residuals fairly constant over the entire longitude range.
The most
conspicuous residuals are seen at $60^\circ \leq l \leq 160^\circ$ and
$-100^\circ \leq l \leq -20^\circ$.
The spiral arm tangents are fairly obvious
in the models and coincide with corresponding peaks in the data profiles
reasonably well indicating that the locations of the spiral arms in the model
are mostly correct.
Therefore, some of the discrepancies observed in the
$l$-$V$ diagrams in Figure~\ref{fig:lvdiagram} are likely due to streaming
motions of the gas that are unaccounted for in the models and can give rise to velocity discrepancies of more than 10~km~s$^{-1}$.
It is also likely that the assumption made in this paper of azimuthally independent radial distribution is not appropriate \citep{KalberlaDedes:2008}.
The residuals in the outer Galaxy for longitudes in the range $90^\circ \leq l \leq 180^\circ$ are larger than those for $-180^\circ \leq l \leq -90^\circ$ for both the \hi{} and CO models.
This indicates that the models are missing some structure in that region, possibly related to the Perseus arm.

The CO model is more strongly affected due to the clumpy nature of the CO emission visible in the
$l$-$V$ diagram in Figure~\ref{fig:lvdiagram}.
About $\sim50$\% of the CO
intensity for $|l| \leq 45^\circ$ is unaccounted for by the model, and there
are positive residuals around $l \pm 90^\circ$.
The very bright peak in CO
emission toward the GC is also not reproduced by the bulge/bar structure.
Outside these longitudes the CO model fares better mostly because the CO
intensity is low toward the outer Galaxy.  There is also the assumption
of continuity for the spiral arms. It is not guaranteed that the spiral arms
all follow a logarithmic shape throughout their distance. The arms are also
assumed to all follow the same radial distribution in density which may not be
the case. This has the effect that even though the
fraction of the density contained in the spiral arms is higher in the best-fit CO
model compared to the best-fit \hi{} model, the peaks associated with the spiral arm
tangents are less prominent in the CO model. There is clear lack of emission around the tangents of the
Perseus arm at $l\approx -23^\circ$ and the Sagittarius-Carina arm at
$l\approx 51^\circ$ and $l\approx -79^\circ$. The CO spatial distribution model
developed is the initial attempt at decomposing the emission into spiral
arms and disk components, given the methodology employed in this paper.
Additional model tuning is required to improve the representation of the
data, but this is deferred to future work.

\section{Effects on CR propagation and high-energy \gray{} emission}
\label{sec:CRgamma}

\subsection{The GALPROP code}
\label{sec:GALPROP}

The underlying concept of the \GP{} code is that all CR-related data,
including direct measurements and indirect electromagnetic observations are
related to the same Milky Way galaxy and should, therefore, be modeled
self-consistently.
With over 20 years of development it is the best known and
most feature-rich code for calculations of CR propagation and their
interactions in the Galactic ISM.
The recently released \GP{}
code version 56 \citep{PorterEtAl:2017} is used in this work.
The latest releases are always
available at the dedicated website\textsuperscript{\ref{fn:link}}, which also provides the WebRun facility to run \GP{} via a web browser interface.
The website contains detailed information on CR propagation together with links to all \GP{} team publications and the supporting data sets required to run the code.
The interstellar gas models developed in this paper will also become available at the above-mentioned website.
A brief overview of the \GP{} code is given below, while further details can
be found in recent \GP{} publications \citep[e.g.,][and references therein]{PorterEtAl:2017,JohannessonEtAl:2016,VladimirovEtAl:2011}.

The \GP{} code numerically solves the diffusion-reacceleration equation for CR
transport with a given source distribution and boundary conditions for all CR
isotopes.
Energy losses from ionization and Coulomb interactions are included
for all species and for CR electrons and positrons energy losses due to
Bremsstrahlung, inverse Compton (IC) scattering, and synchrotron emission are
also included.
Additional processes for nuclei include nuclear spallation,
secondary particle production, radioactive decay, electron capture and
stripping, electron knock-on, and electron K-capture.
Electromagnetic
radiation from the decays of $\pi^0$, $K^0$, and heavier mesons,
Bremsstrahlung, IC scattering, and synchrotron emission are calculated
self-consistently once the system of transport equations has been solved.
To capture the finer structure of the interstellar gas the \gray\ intensity maps associated with interactions between CRs and
the interstellar gas use the column density estimated from the line
emission surveys that have been split into Galactocentric annular maps using the gas velocity
field; see Appendix~B of \citet{AckermannEtAl:2012} for full details of their construction.
The \GP{}
code has proven to be remarkably successful in modeling both CR data
\citep[e.g.,][]{JohannessonEtAl:2016,BoschiniEtAl:2017} and electromagnetic
radiation associated with CR interactions in the ISM
\citep[e.g.,][]{AckermannEtAl:2012,AjelloEtAl:2016}.

The spatial number density distribution of the interstellar gas is used in
\GP{} for calculations of the production of secondaries and energy losses by
CR species.
In addition, the number density distribution is used internally
for the proper weighting of the gas column density along the LOS in the individual
Galactocentric gas rings for calculations of Bremsstahlung and $\pi^0$-decay
\gray{} intensity maps.
However, the density distribution is approximate and
does not represent all details of the true distribution of the interstellar
gas.
Therefore, the ratios between the column densities estimated from the
\hi{} and CO line emission surveys and corresponding column densities from the
Galactic gas distributions employed in \GP{} are used as multiplicative
corrections for each LOS integration.
This method enables \GP{} to predict the
details in the structure of the gas-related interstellar emissions imprinted
in the \gray{} skymaps without the necessity to develop precise 3D models of the
gas distribution in the Galaxy.
Even though all versions of \GP{} since the
very beginning allowed full 3D functionality, including 3D gas distributions,
the analytic 2D cylindrically symmetric gas distributions as described by
\citet{AckermannEtAl:2012} have been most commonly used.
This usage mode is attributed to the increased demand for computing resources
required to run \GP{} in 3D mode as well as the absence of detailed 3D models
of the ISM.
The latest \GP{} (version 56) expands upon this functionality with the capability to read in the same XML description of the gas distribution used in the \GG\ code, and the optimizations made significantly improve the performance.
The output from the analysis described in Section~\ref{sec:gasModeling} can, therefore, be used directly without any modification.

\subsection{CR propagation}
\label{sec:crcalc}

This Section illustrates how the new 3D interstellar gas distributions affect
the CR propagation parameters derived from a fit to the direct observations of
primary and secondary CR species. This study is limited to models with diffusive reacceleration and an isotropic
and homogeneous diffusion coefficient with a power-law rigidity dependence.
The propagation parameters are tuned using the method described by
\citet{PorterEtAl:2017}.
To enable comparison with that work the same CR source density models, SA0, SA50, and SA100 are used.
The CR source density distribution is composed of two constituents, a disk and 4 spiral arms, where each component has the same exponential scale height of 200 pc perpendicular to the Galactic plane.
The source distribution in the disk follows the radial distribution of pulsars as given by \citet{YusifovKucuk:2004}.
The source distribution in the spiral arms matches the density distribution of the stellar population in the four main arms described by \citet{RobitailleEtAl:2012}.
The different numbers $nn$ in the source model name (SA$nn$) represent the percentile contribution from the sources in the spiral arms. 
The spiral arms in the CR source models are not identical to those in the
  gas density model. Due to different pitch angles, some parts of the spiral
  arms in the CR source models end in inter-arm regions of the gas density
  models and vice-versa. While this configuration may not be completely physical, there are
  theories that predict an offset between the peak of the gas distribution and
  that of star formation that should trace the CR source distribution
  \citep{Vallee:2014}. Having a single model both with and without an offset
  enables an illustration of the effects of both models within a single
calculation.

Each CR source model is then paired with the standard 2D interstellar gas distributions available in \GP{} \citep{AckermannEtAl:2012} and the new 3D gas distributions described in this paper, resulting in a total of 6 models.
To identify changes associated with the choice of the gas distribution, the same standard 2D ISRF model is used for all 6 \GP{} models \citep{AckermannEtAl:2012}.
The SA0--2D gas model is used as the reference case for comparison with the other models considered in this paper.
This combination corresponds to the 2D CR source and gas distributions scenario that has been the standard approach for CR and interstellar emission modeling in the past and is the same reference model used by \citet{PorterEtAl:2017}.

The main effects of varying the interstellar gas distributions are expected at low energies where the interstellar CR propagation is slow and the energy losses are fast.
A comparison of propagated CR spectra with the direct measurements made deep in the heliosphere is impossible without taking into account heliospheric effects. Because the details of the heliospheric propagation and the CR spectra depend on the solar activity, the effect of variation of CR fluxes is called heliospheric or solar modulation.
The modulated spectra of CR species differ considerably from the local interstellar spectra below $\sim$20-50 GeV/nucleon, where the effect becomes stronger as energy decreases. 

\begin{deluxetable}{lcc}[tb!]
\tablecolumns{3}
\tablewidth{0pc}
\tablecaption{CR data used for determination of the propagation parameters
 \label{tab:CRdata}}
\tablehead{
\colhead{Instruments} &
\colhead{Species} &
\colhead{References\tablenotemark{a}}
}
\startdata
AMS-02 (2011-2016) & B/C & [1] \\
AMS-02 (2011-2013) & $e^-$ & [2] \\
AMS-02 (2011-2013) & H & [3] \\
AMS-02 (2011-2013) & He & [4] \\
HEAO3-C2 (1979-1980) & B, C, O, Ne, Mg, Si & [5] \\
Voyager-1 (2012-2015) & H, He, B, C, O, Ne, Mg, Si & [6] \\
PAMELA (2006-2008) & B, C & [7]
\enddata
\tablenotetext{a}{[1] \citet{PRL:1171102}, [2] \citet{PRL:1131102}, [3] \citet{PRL:1141103}, [4] \citet{PRL:1151101}, [5] \citet{Engelmann:1990}, [6] \citet{CummingsEtAl:2016}, [7] \citet{AdrianiEtAl:2014}.}
\end{deluxetable}

\begin{deluxetable*}{@{\extracolsep{4pt}}lcccccc@{}}[th!]
\tablecolumns{7}
\tablewidth{0pc}
\tablecaption{Final \GP{} model parameters \label{tab:CRparameters} }
\tablehead{
 & \multicolumn{3}{c}{2D gas models} & \multicolumn{3}{c}{3D gas
 models} \\
\cline{2-4}\cline{5-7}\noalign{\rule{0pt}{.5ex}}
\colhead{Parameter} & 
\colhead{SA0} & 
\colhead{SA50} &
\colhead{SA100} &
\colhead{SA0} & 
\colhead{SA50} &
\colhead{SA100} 
}
\startdata
\tablenotemark{a }$D_{0}$, $10^{28}$ cm$^2$ s$^{-1}$ & $4.37$ & $4.47$ & $4.71$ & $2.20$ & $2.28$ & $2.34$ \\

\tablenotemark{a }$\delta$ & $0.494$ & $0.508$ & $0.483$ & $0.546$ & $0.545$ & $0.549$ \\

\phantom{\tablenotemark{a }}$v_{A}$, km s$^{-1}$ & $7.64$ & $9.19$ & $7.34$ & $5.86$ & $5.26$ & $3.97$ \\

\tablenotemark{b }$\gamma_0$ & $1.47$ & $1.61$ & $1.66$ & $1.37$ & $1.51$ & $1.51$ \\

\tablenotemark{b }$\gamma_1$ & $2.366$ & $2.350$ & $2.381$ & $2.338$ & $2.345$ & $2.357$ \\

\tablenotemark{b }$\rho_1$, GV & $3.64$ & $3.92$ & $4.12$ & $3.40$ & $3.56$ & $3.33$ \\

\tablenotemark{b }$\gamma_{0,\text{H}}$ & $1.75$ & $1.77$ & $1.78$ & $1.75$ & $1.71$ & $1.79$ \\

\tablenotemark{b }$\gamma_{1,\text{H}}$ & $2.375$ & $2.359$ & $2.349$ & $2.331$ & $2.349$ & $2.322$ \\

\tablenotemark{b }$\gamma_{2,\text{H}}$ & $2.199$ & $2.200$ & $2.238$ & $2.203$ & $2.190$ & $2.219$ \\

\tablenotemark{b }$\rho_{1,\text{H}}$, GV & $5.99$ & $5.99$ & $5.67$ & $5.32$ & $4.81$ & $4.93$ \\

\tablenotemark{b }$\rho_{2,\text{H}}$, GV & $265$ & $225$ & $403$ & $206$ & $200$ & $206$ \\

\phantom{\tablenotemark{a }}$\Delta_{\text{He}}$ & $0.034$ & $0.034$ & $0.039$ & $0.043$ & $0.045$ & $0.035$ \\

\tablenotemark{b }$\gamma_{0,e}$ & $1.66$ & $1.67$ & $1.57$ & $1.63$ & $1.81$ & $1.74$ \\

\tablenotemark{b }$\gamma_{1,e}$ & $2.761$ & $2.753$ & $2.749$ & $2.744$ & $2.769$ & $2.734$ \\

\tablenotemark{b }$\gamma_{2,e}$ & $2.351$ & $2.327$ & $2.312$ & $2.305$ & $2.378$ & $2.303$ \\

\tablenotemark{b }$\rho_{1,e}$, GV & $5.82$ & $5.89$ & $6.14$ & $5.68$ & $5.97$ & $6.90$ \\

\tablenotemark{b }$\rho_{2,e}$, GV & $102$ & $101$ & $102$ & $100$ & $76$ & $109$ \\

\tablenotemark{c }$J_\text{H}$, $10^{-9}$ cm$^{-2}$ s$^{-1}$ sr$^{-1}$ MeV$^{-1}$ & $4.520$ & $4.498$ & $4.610$ & $4.486$ & $4.542$ & $4.322$ \\

\tablenotemark{c }$J_e$, $10^{-11}$ cm$^{-2}$ s$^{-1}$ sr$^{-1}$ MeV$^{-1}$ & $1.242$ & $1.252$ & $1.243$ & $1.290$ & $1.316$ & $1.231$ \\

\tablenotemark{d }$q_{0,^{4}\text{He}}/q_{0,\text{H}}\times10^{6}$ & $ 94602$ & $ 95324$ & $ 97365$ & $101800$ & $100160$ & $100630$ \\
 
\tablenotemark{d }$q_{0,^{12}\text{C}}/q_{0,\text{H}}\times10^{6}$ & $ 2882$ & $ 2867$ & $ 2746$ & $ 2960$ & $ 2916$ & $ 2849$ \\

\tablenotemark{d }$q_{0,^{16}\text{O}}/q_{0,\text{H}}\times10^{6}$ & $ 3780$ & $ 3873$ & $ 3645$ & $ 3944$ & $ 3950$ & $ 3804$ \\

\tablenotemark{d }$q_{0,^{20}\text{Ne}}/q_{0,\text{H}}\times10^{6}$ & $ 356$ & $ 358$ & $ 333$ & $ 379$ & $ 371$ & $ 356$ \\

\tablenotemark{d }$q_{0,^{24}\text{Mg}}/q_{0,\text{H}}\times10^{6}$ & $ 644$ & $ 654$ & $ 609$ & $ 685$ & $ 675$ & $ 657$ \\

\tablenotemark{d }$q_{0,^{28}\text{Si}}/q_{0,\text{H}}\times10^{6}$ & $ 742$ & $ 762$ & $ 718$ & $ 779$ & $ 783$ & $ 756$ \\

\tablenotemark{e }$\Phi_{\text{HEAO3-C2}}$, MV & $ 857$ & $ 849$ & $ 827$ & $ 845$ & $ 850$ & $ 833$ \\

\tablenotemark{e }$\Phi_{\text{PAMELA}}$, MV & $ 578$ & $ 578$ & $ 572$ & $ 584$ & $ 587$ & $ 582$ \\

\tablenotemark{e }$\Phi_{\text{AMS}}$, MV & $ 638$ & $ 645$ & $ 581$ & $ 729$ & $ 768$ & $ 649$ 
\enddata
\tablenotetext{a}{$D(\rho) \propto \beta \rho^{\delta}$ where $\rho$ is the rigidity. $D(\rho)$ is normalized to $D_0$ at 4~GV.}
\tablenotetext{b}{The injection spectrum is parameterized as $q(\rho) \propto \rho^{\gamma_0}$ for $\rho < \rho_1$, $q(\rho) \propto \rho^{\gamma_1}$ for $\rho_1 < \rho < \rho_2$, and $q(r) \propto \rho^{\gamma_2}$ for $\rho > \rho_2$. The spectral shape of the injection spectrum is the same for all species except H and He.}
\tablenotetext{c}{The proton and $e^-$ fluxes are normalized at the Solar location at the kinetic energy of 100~GeV.}
\tablenotetext{d}{The injection spectra for isotopes are adjusted as a ratio of the proton injection spectrum at 100~GeV. The isotopes not listed here have the same value as found in \citet{JohannessonEtAl:2016}.}
\tablenotetext{e}{Solar modulation is calculated using the force-field
  approximation. }
\end{deluxetable*}

CR propagation in the heliosphere is described by the \citet{1965P&SS...13....9P} equation.
Spatial diffusion, convection with the solar wind, drifts, and adiabatic cooling are the main processes that influence transport of CRs to the inner heliosphere.
These effects have been incorporated into realistic (time-dependent, 3D) models \citep[e.g.,][]{2003JGRA..108.1228F,2006ApJ...640.1119L,2004AnGeo..22.3729P,BoschiniEtAl:2017}.
There is considerable degeneracy between the parameters of the heliospheric propagation models and those controlling the low-energy behavior of the Galactic CR propagation \citep{BoschiniEtAl:2017}.
So far, the detailed analysis was made only for CR protons, helium, antiprotons,
and electrons \citep{BoschiniEtAl:2017, BoschiniEtAl:2018}.
Evaluation of CR propagation parameters involves propagation of elements
heavier than oxygen (Table~\ref{tab:CRdata}), for which the same thorough analysis is not
available yet. Besides, the current work is aimed at the study of the effects
of different gas distributions on the interstellar CR propagation. Therefore,
the simplest available heliospheric modulation is used, the so-called ``force-field'' approximation
\citep{1968ApJ...154.1011G}
It characterizes the whole complexity of the time-dependent heliospheric modulation with a single parameter -- the ``modulation potential''.
Such an approach has no predictive power, but has been widely used as a simple low-energy parameterization of the modulated spectrum. 

The interstellar propagation parameters are tuned using a maximum-likelihood
fit employing the data sets listed in Table~\ref{tab:CRdata}.
To reduce the number of free parameters in each fit, the procedure is split into two stages, similar to the analysis described in \citet{CummingsEtAl:2016}.
The propagation model parameters that are fit for are listed in Table~\ref{tab:CRparameters}.
There is a strong degeneracy between the halo height and the normalization of the diffusion coefficient.
Even though using the radioactive-clock isotopes ($^{10}$Be, $^{26}$Al, $^{36}$Cl, $^{54}$Mn) constrains the halo size significantly, the range of possible values remains quite large \citep{JohannessonEtAl:2016}.
Instead of fitting for both, the diffusion coefficient and the halo size, simultaneously, the halo height is fixed to 6~kpc, in good agreement with previous analyses \citep[e.g.,][]{MoskalenkoEtAl:2005,OrlandoStrong:2013,JohannessonEtAl:2016}.

At the first stage, the interstellar propagation parameters
are fitted together with the injection spectra and abundances of
elements heavier than helium.
With the propagation parameters and the injection
spectra for those elements determined, they are held constant.
The injection
spectra for electrons, protons, and helium are then obtained at the second stage.
To reduce the number of parameters the injection spectrum of helium is coupled
with that of protons so that the breaks are at the same rigidities, and
spectral indices of helium are smaller than the protons by a parameter
$\Delta_{\text{He}}$ that is also derived from the fit.
This is similar to the linking the proton and helium spectra in the analysis described in \citet{JohannessonEtAl:2016}.
Fourteen parameters are determined at the first stage of the procedure, while the second stage fits for fifteen parameter values.

The calculations are made for a Cartesian spatial grid with dimensions $\pm 20$~kpc for the $X$ and $Y$ coordinates, with $\Delta X=\Delta Y = 0.2$~kpc, $\Delta Z = 0.1$~kpc and CR kinetic energy grid covering 10~MeV/nucleon to 100~TeV/nucleon with logarithmic spacing at 16 bins/decade.
The span and sampling of the spatial and energy grids is chosen to enable realistic and efficient computations given the available resources\footnote{Increasing the energy grid sampling by a factor of 2 only produces a change in the propagated CR intensities at maximum of $\sim$$2$\%. The runtime and memory consumption is increased by a proportional factor for the finer energy grid, but would not substantially alter the results or conclusions.}.
The spatial grid sub-division size allows adequate sampling of the CR and ISM density distributions.
The $X,Y$ size of the grid is sufficient to ensure that CR leakage from the Galaxy is determined by the halo height rather than the extent of the $X,Y$ grid.
It has been shown that there is only a weak effect on parameters determined for 2D models using 20~kpc and 30~kpc radial boundaries even with halo heights as large as 10~kpc \citep{AckermannEtAl:2012}.

\begin{figure*}[t]
\center{
\includegraphics[width=0.50\textwidth]{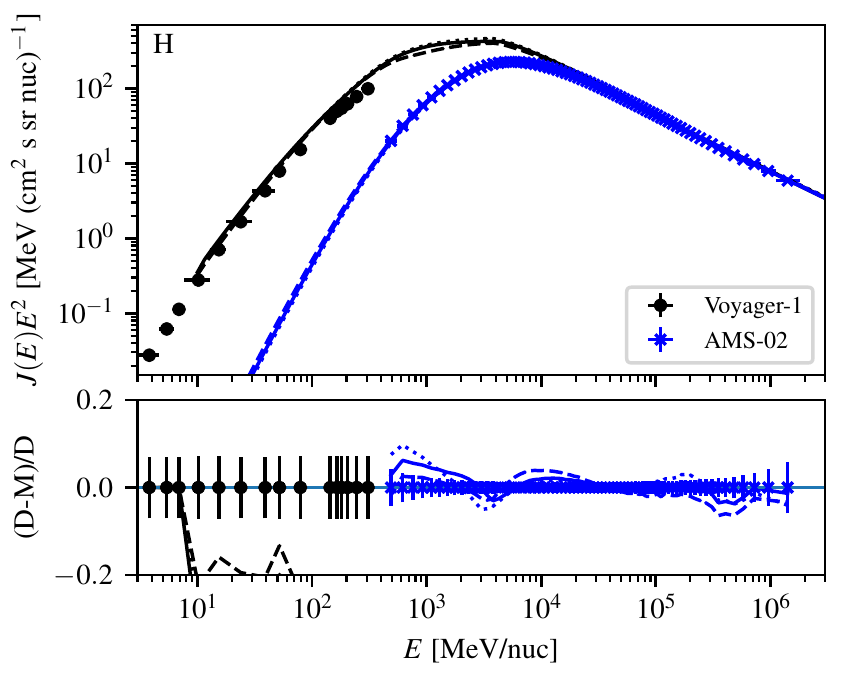}\hfill
\includegraphics[width=0.50\textwidth]{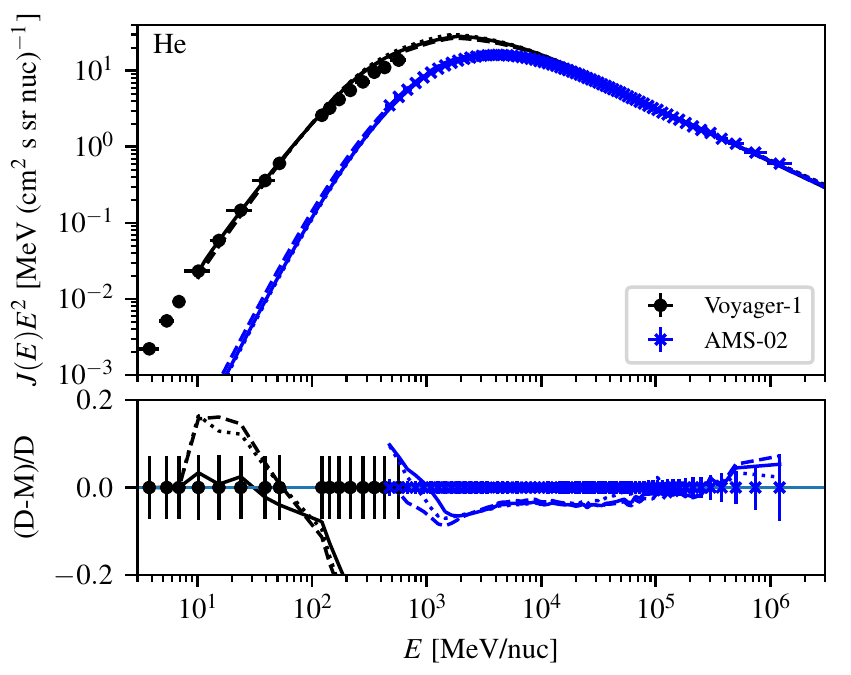}\\
\includegraphics[width=0.50\textwidth]{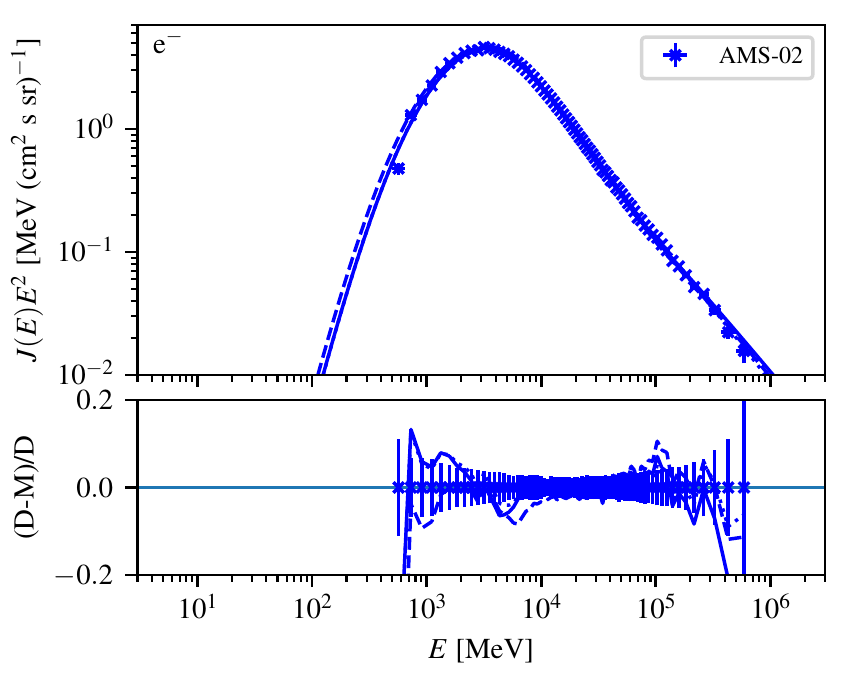}\hfill
\includegraphics[width=0.50\textwidth]{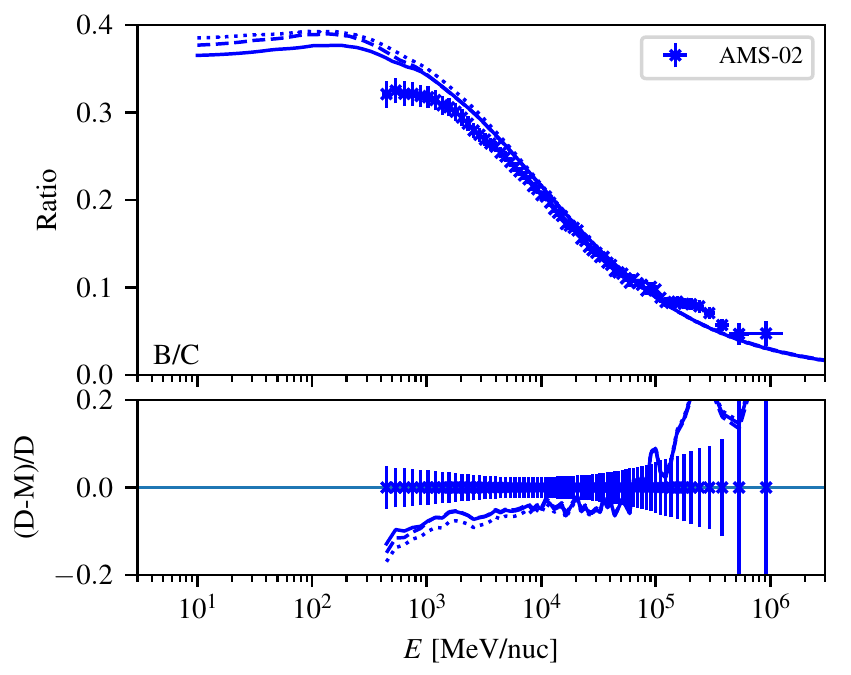}\\
\includegraphics[width=0.50\textwidth]{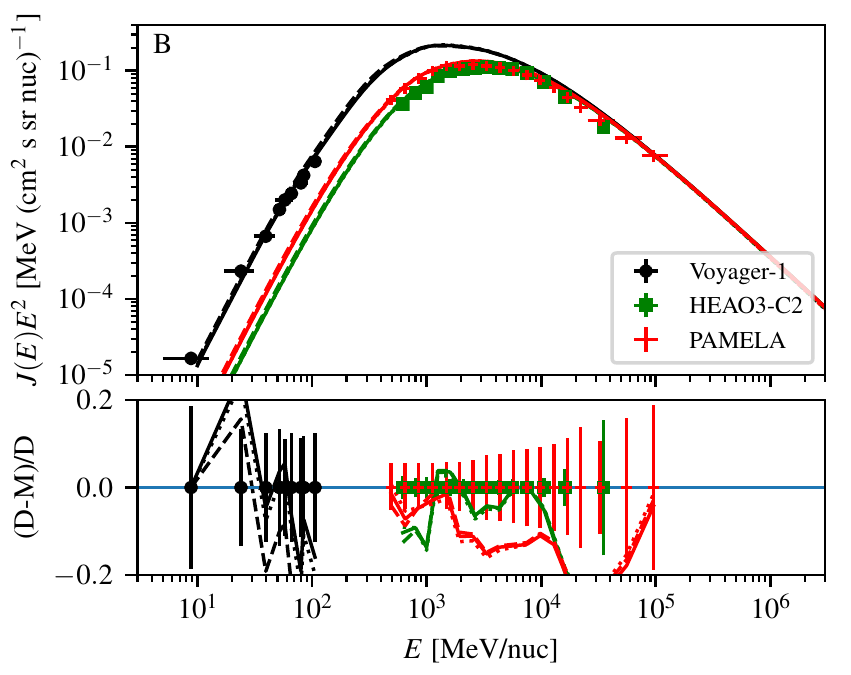}\hfill
\includegraphics[width=0.50\textwidth]{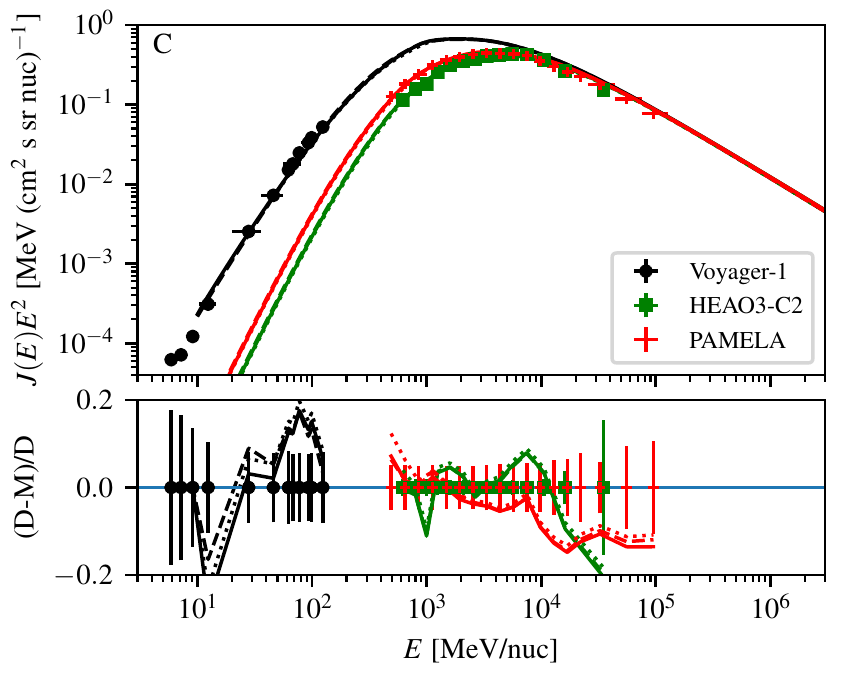}
}
\caption{\GP{} predictions calculated using the new 3D gas distributions and
CR source density models compared with CR data: SA0 (solid curve), SA50
(dotted curve), and SA100 (dashed curve). Shown are CR species: protons (top
left), helium (top right), $e^-$ (center left),  boron over carbon ratio (center right), boron (bottom left), and carbon (bottom right). Bottom panel of each figure shows the fractional residuals. }
\label{fig:Spectra1}
\end{figure*}

Table~\ref{tab:CRparameters}\footnote{The parameters for the 2D gas model are reproduced from \citet{PorterEtAl:2017}} shows the results of the fitting procedure for the SA0, SA50, and SA100 CR source density models for both 2D and 3D gas models.
The model predictions for the CR flux at the location of the Sun are very
similar being within $\sim$$5$\% of each other as shown in Figure~\ref{fig:Spectra1}.
The latter is not surprising because the models are fit to the CR data.
All models generally agree with data better than 10\% with deviations reaching up to 20\% for some energy ranges and elements.
This level of agreement with data is sufficient for the purpose of this paper.

\begin{figure*}[tb!]
\center{
\includegraphics[width=0.33\textwidth]{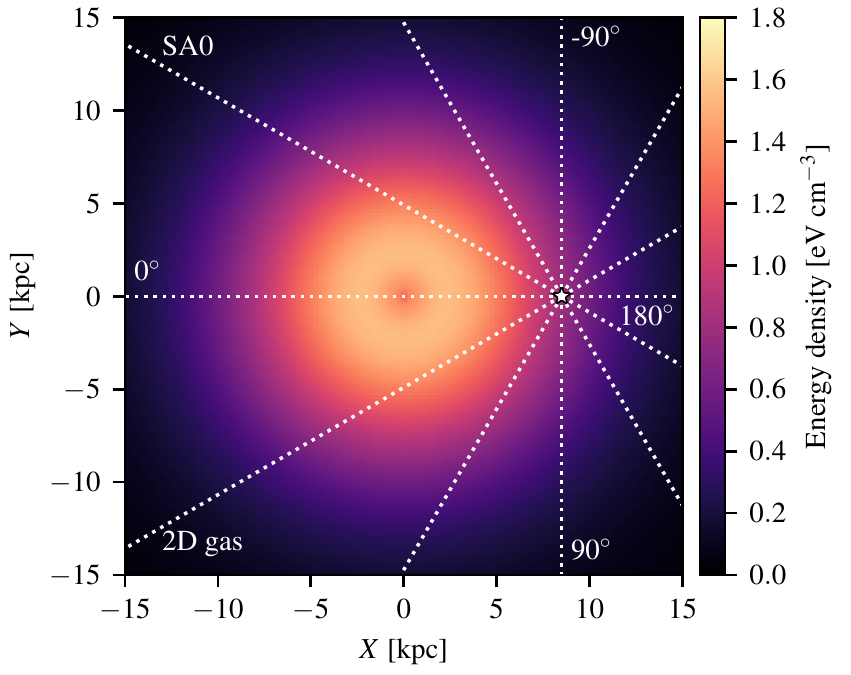}\hfill
\includegraphics[width=0.33\textwidth]{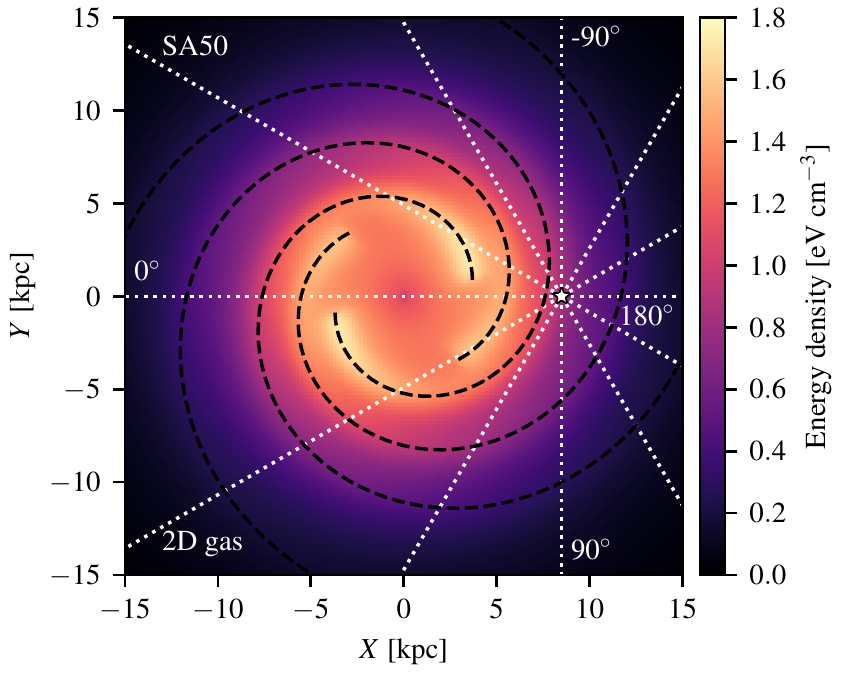}\hfill
\includegraphics[width=0.33\textwidth]{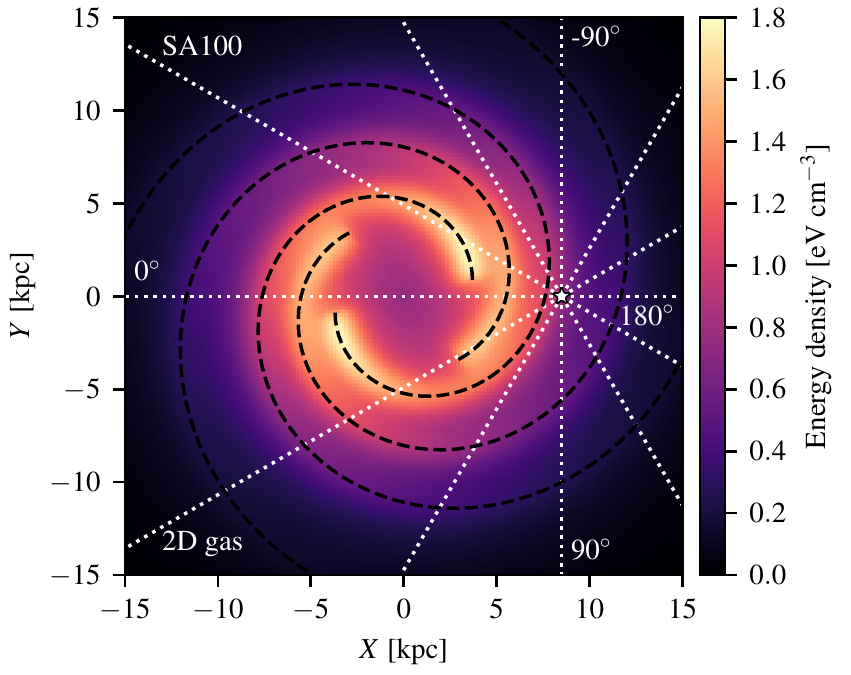}\\
\includegraphics[width=0.33\textwidth]{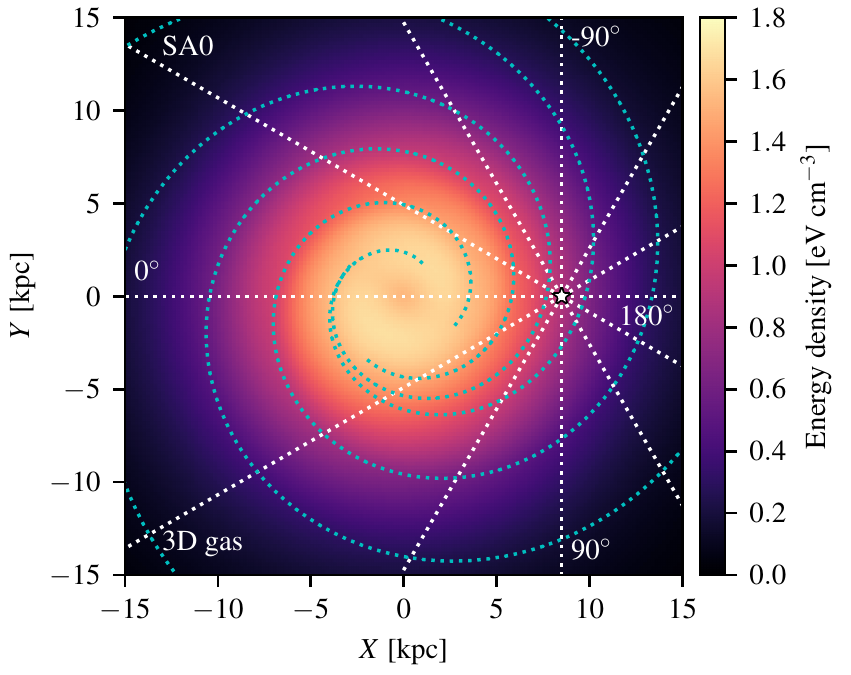}\hfill
\includegraphics[width=0.33\textwidth]{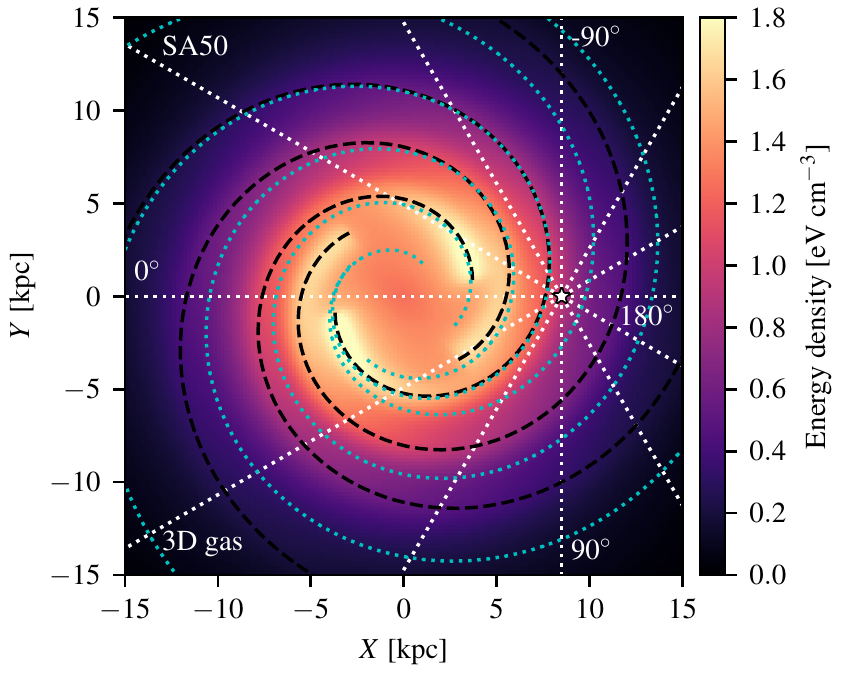}\hfill
\includegraphics[width=0.33\textwidth]{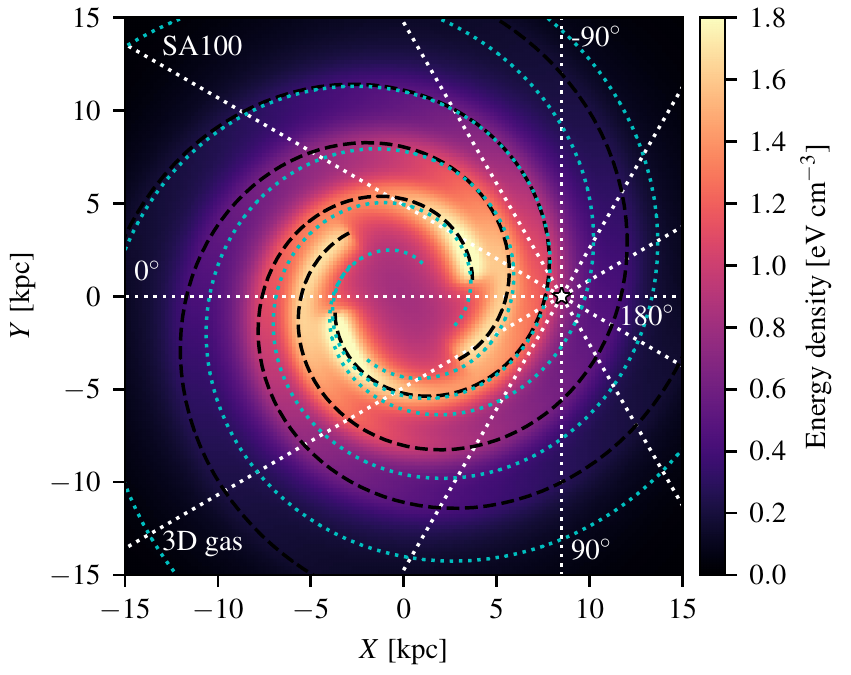}
}
\caption{Total CR energy density in the mid-plane of the Galaxy ($Z=0$) for SA0, SA50, and SA100 CR source distributions (left to right, respectively) with the 2D gas distribution on top and the 3D gas distribution at the bottom. The solar location is marked as a white point and the white dashed lines mark the longitude grid with 30$^\circ$ step. The black dashed curves trace the positions of the spiral arms in the CR source distribution and the cyan dotted curves trace the spiral arms in the 3D gas distribution.}
\label{fig:CRdensity}
\end{figure*}

Comparison of the best-fit parameters of the models employing the 2D gas
distributions with those using the new 3D gas distributions shows that
there are significant changes in the normalization of the diffusion
coefficient $D_0$ and its rigidity dependence $\delta$.
There are also corresponding changes in the Alfv\'en velocity values $v_A$.
Using the 3D gas distributions results in slower diffusion and weaker
diffusive reacceleration.
The main reason for this change in the parameter values is the lower local gas density for the 3D gas models.
The average surface density of the combined interstellar gas within a few kpc of the Sun using the new 3D distributions is nearly a factor of 2 lower than that calculated using the 2D distributions.
This is the region where most of the secondary Boron reaching the Earth is produced \citep{JohannessonEtAl:2016}.
The surface density, rather than central plane number density, is used  for inter-comparison because the scale height of the CR diffusion zone is significantly larger than that of the gas and the surface density thus provides better estimate of total column density of the gas traversed by CRs.
The difference in total surface density of about a factor of 2 agrees reasonably well with the factor of 2 change in $D_0$.
In turn, slower diffusion results in weaker diffusive acceleration that is needed to reproduce the data.

The reason for this significant change in the local gas surface density
between the distributions is two-fold.
First the method for determination of
the gas distribution using a student-t likelihood favors models that
under-predict the data leaving residuals that can be absorbed by additional
model components.
This is evident from the positive residuals seen in
Figure~\ref{fig:longitudeProfiles}.
Those cannot, however, explain a factor of 2 difference because the residuals around $l \sim 90^\circ$ and $l \sim-90^\circ$ are $\sim 40$\% of the model and
should therefore account for only about half of the difference.
Another important difference between the gas distributions is
that the older 2D distributions were
derived assuming the Sun is located at 10~kpc from the GC
\citep{GordonBurton:1976, BronfmanEtAl:1988}.
These distributions have not been scaled to the IAU recommended value of $R_\odot=8.5$~kpc that is used for the \GP{} calculations.
Convolving the old 2D gas distributions with \GG{} using the
updated rotation curve and the updated solar location results in a significant over-prediction of the data.
The 2D gas distributions for the newly
released \GP{} version 56 have been re-scaled to the correct Sun-GC distance
of $R_\odot=8.5$~kpc.
The scaled 2D gas distributions provide a good agreement with the \hi\
and CO data.
Using the rescaled 2D gas distributions in \GP\ in a fit to the CR data
results in $D_0 = 3.16\times10^{28}$~cm$^{2}$s$^{-1}$, which is about half-way between the results for the old 2D distributions and those obtained with the new 3D
distributions.
For compatibility with previous work and, in particular,
the work by \citet{PorterEtAl:2017} the
calculations here do not use the corrected 2D gas distributions.
\GP{} is not the only propagation code
that uses these incorrectly scaled distributions, other
codes that have incorporated the \GP{} analytic gas code and use it with $R_\odot=8.5$ kpc are also susceptible to the error.

The change in propagation parameters between the different CR source density models is small for both 2D and 3D gas distributions, but statistically significant.
There is no obvious trend for most of the parameters.
That is, the values for those of SA50 are not always between the values for SA0 and SA100.
Note that the values of $v_A$ and $\delta$ determined here for the \GP{} 2D gas distributions differ from those obtained by \citet{JohannessonEtAl:2016} and \citet{CummingsEtAl:2016} because of the datasets employed.
The larger value of the delta parameter comes from the reduced Alfv\'{e}n speed obtained from the fits: higher Alfv\'{e}n speeds result in a larger bump around $\sim$$1$ GeV in the B/C ratio than that required by the AMS-02 data used in this paper.

Figure~\ref{fig:CRdensity} shows the total energy density of CRs in the Galactic plane for the 6 models considered in this paper.
It illustrates how the gas and CR source distribution affects the final distribution of CRs in the Galaxy after propagation.
The spiral arm structure of the CR source distribution is clearly visible.
The visible width of the spiral arms is considerably larger than the input CR sources because of the diffusion.
The change in diffusion parameters between the two gas distributions creates a visible and significant effect where the spiral arm features are distinctly sharper with the new gas distributions and smaller diffusion coefficient.
The imprint on the gas distribution can also be seen for the SA0 CR source distribution where the enhanced density in the spiral arm causes faster cooling. 

\begin{figure*}[tb!]
 \centering
 \includegraphics[width=0.33\textwidth]{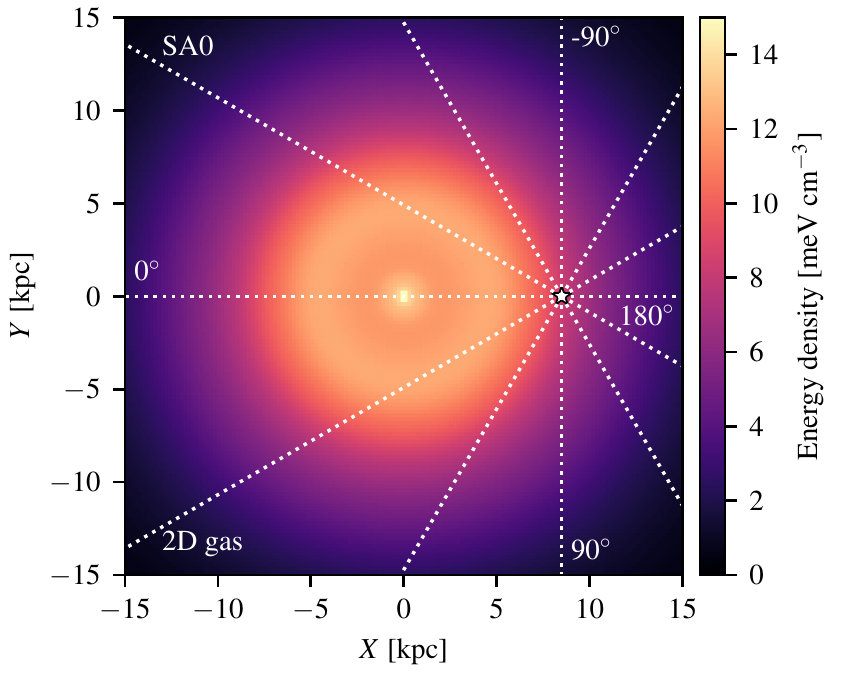}\hfill
 \includegraphics[width=0.33\textwidth]{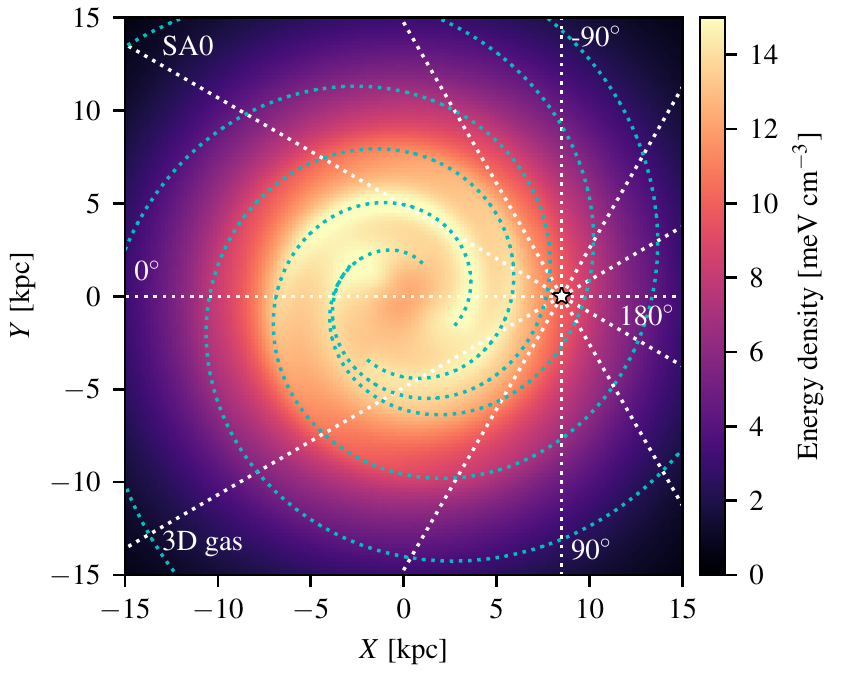}\hfill
 \includegraphics[width=0.33\textwidth]{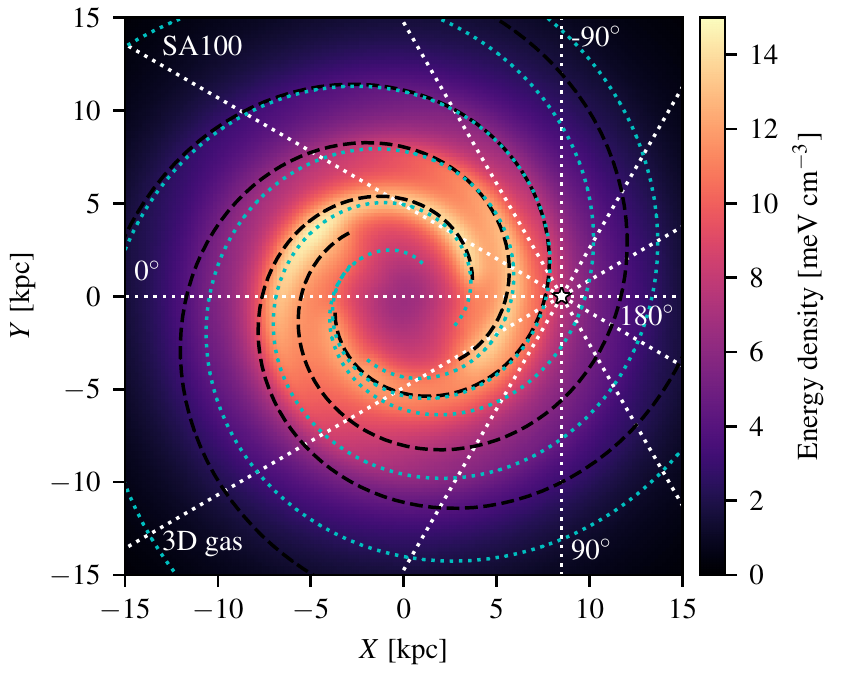}\\
 \includegraphics[width=0.33\textwidth]{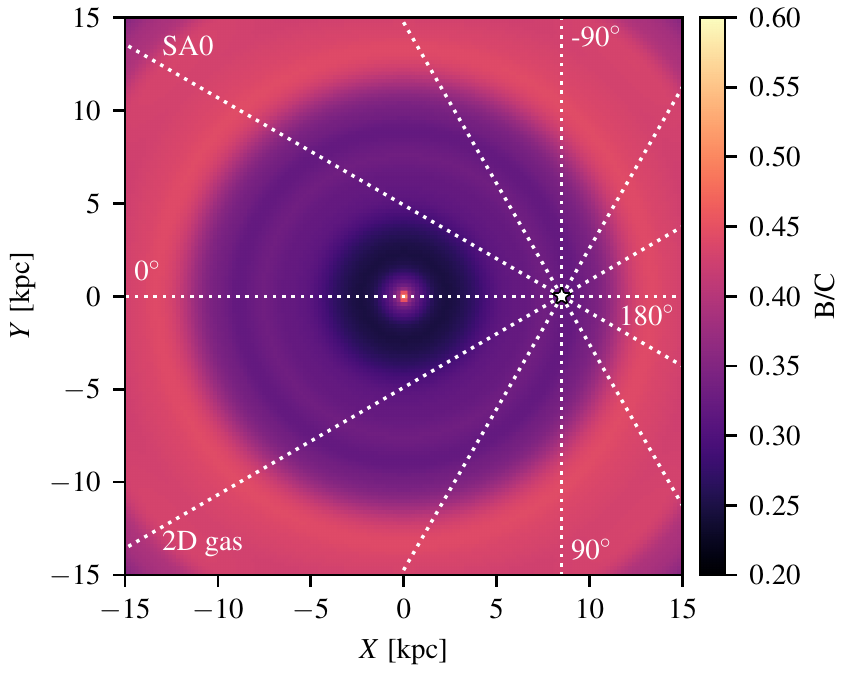}\hfill
 \includegraphics[width=0.33\textwidth]{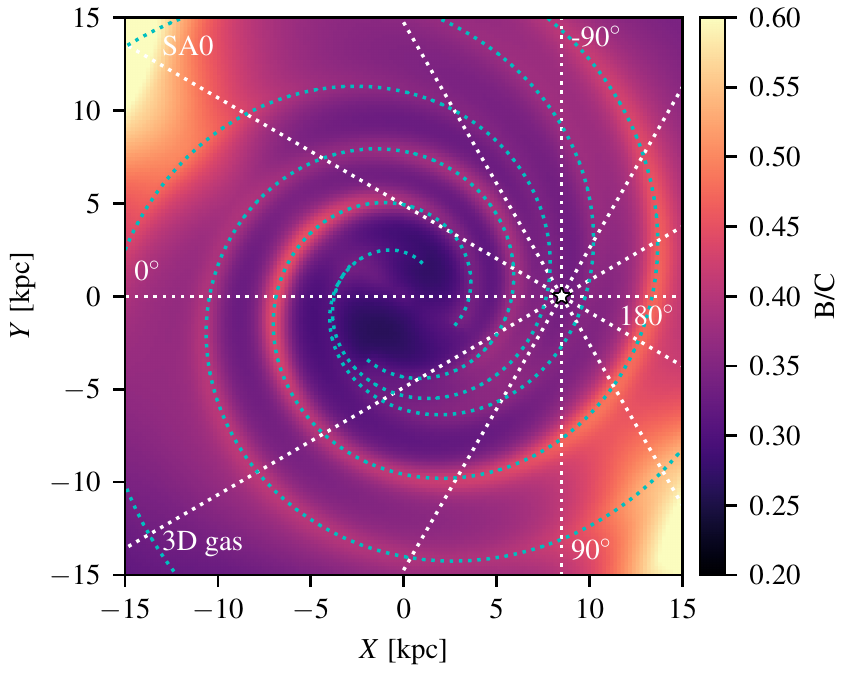}\hfill
 \includegraphics[width=0.33\textwidth]{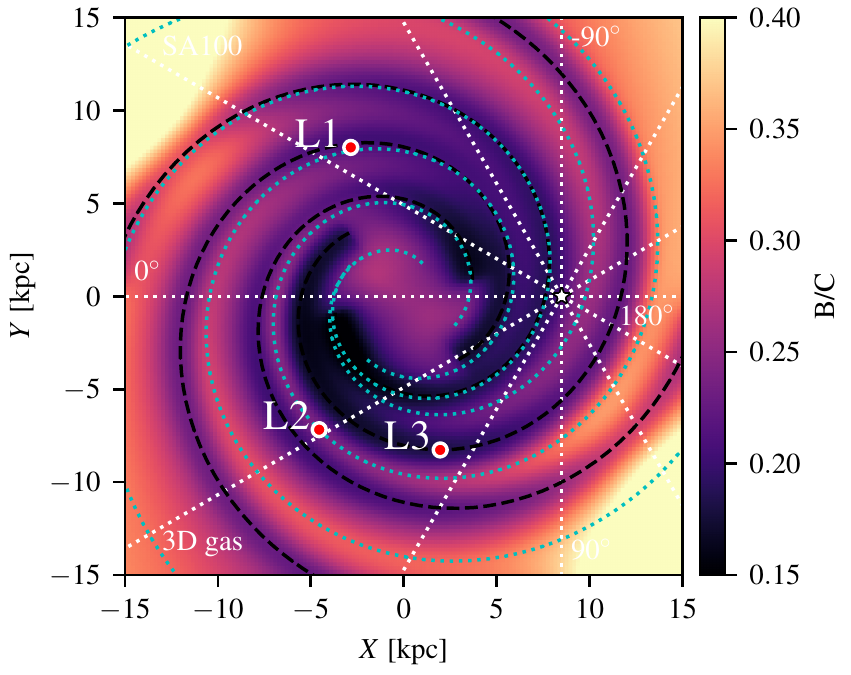}
 \caption{Top row: CR energy density of secondary positrons in the mid-plane
   of the Galaxy for the SA0-2D gas (left), SA0-3D gas (center), and SA100-3D
   gas (right). Bottom row: B/C ratio at 10~GeV/nuc in the mid-plane of the
   Galaxy for same models. The lines and curves are the same as in
   Figure~\ref{fig:CRdensity}. The three labeled red dots with the
   white border in the bottom right
   panel mark locations for the CR spectra shown in Figure~\ref{fig:secondaryRatios}. }
 \label{fig:CRsecondaries}
\end{figure*}

\begin{figure*}[tb!]
 \centering
 \includegraphics[width=0.33\textwidth]{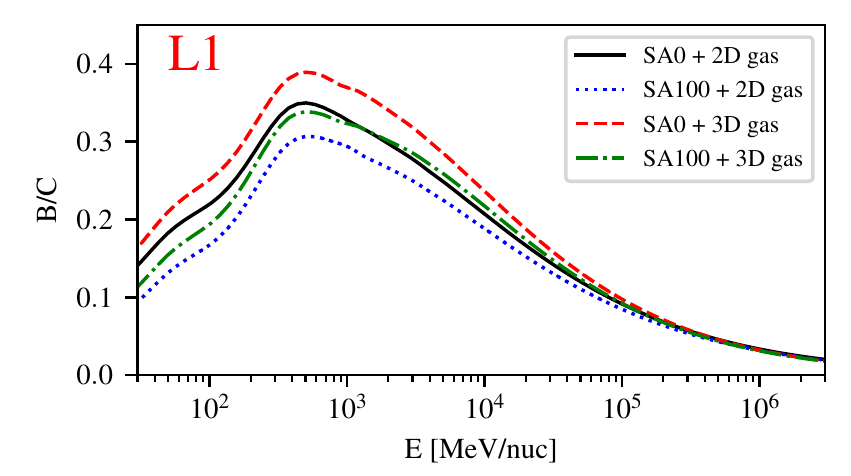}\hfill
 \includegraphics[width=0.33\textwidth]{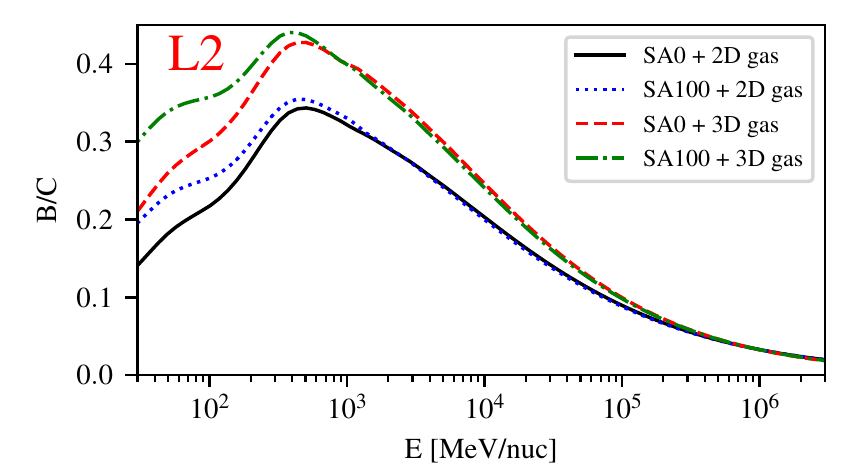}\hfill
 \includegraphics[width=0.33\textwidth]{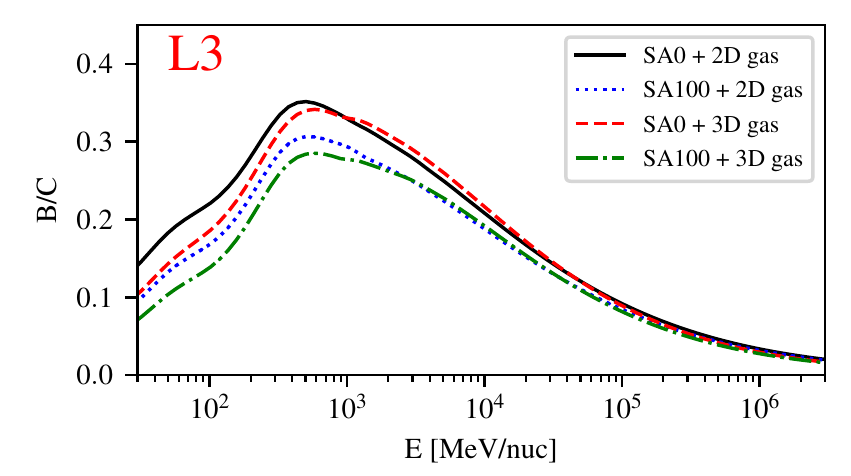}\\
 \includegraphics[width=0.33\textwidth]{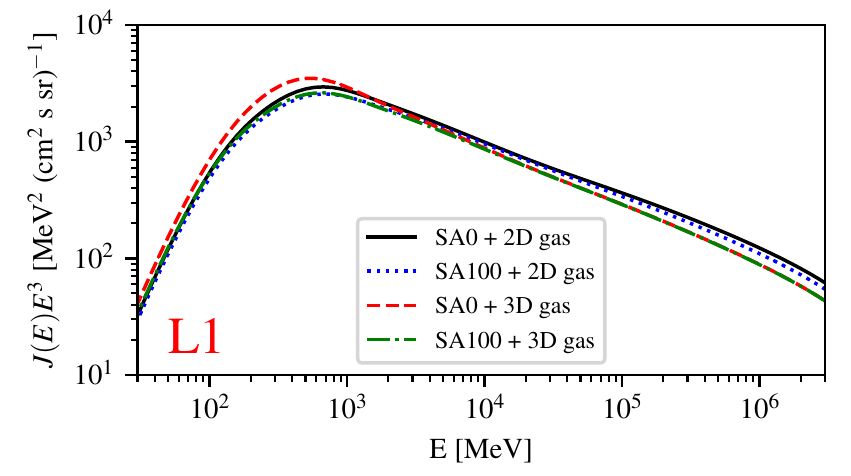}\hfill
 \includegraphics[width=0.33\textwidth]{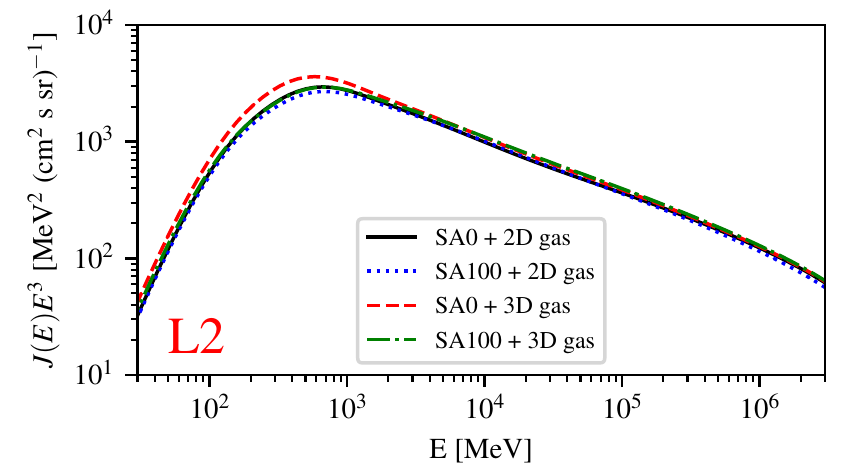}\hfill
 \includegraphics[width=0.33\textwidth]{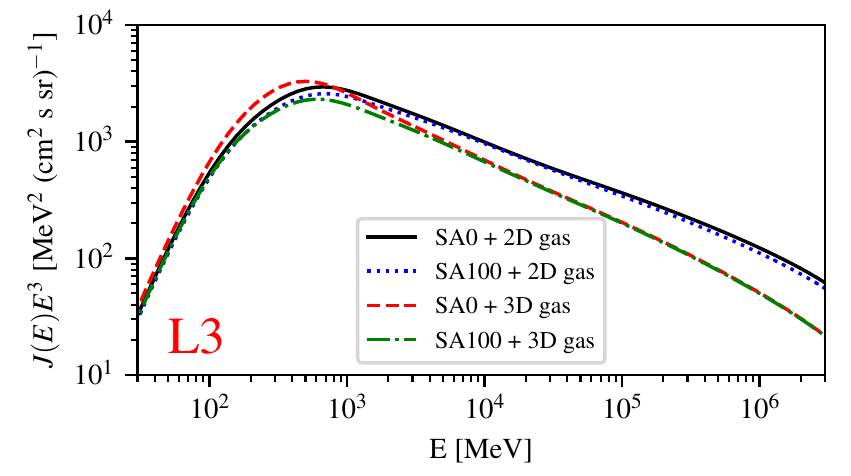}
 \caption{Top row: B/C ratio for SA0-2D (black solid curve), SA100-2D (blue
   dotted curve), SA0-3D (red dashed curve), and SA100-3D (green dash-dotted
   curve) models at the positions marked by the red dots with the white border
   in Figure~\ref{fig:CRsecondaries}. Bottom row: positron spectrum
 plotted at those same locations for the same models. The red label in
 the panels corresponds to the appropriate dot location.}
 \label{fig:secondaryRatios}
\end{figure*}

\begin{figure*}[tb!]
\center{
\includegraphics[width=0.32\textwidth]{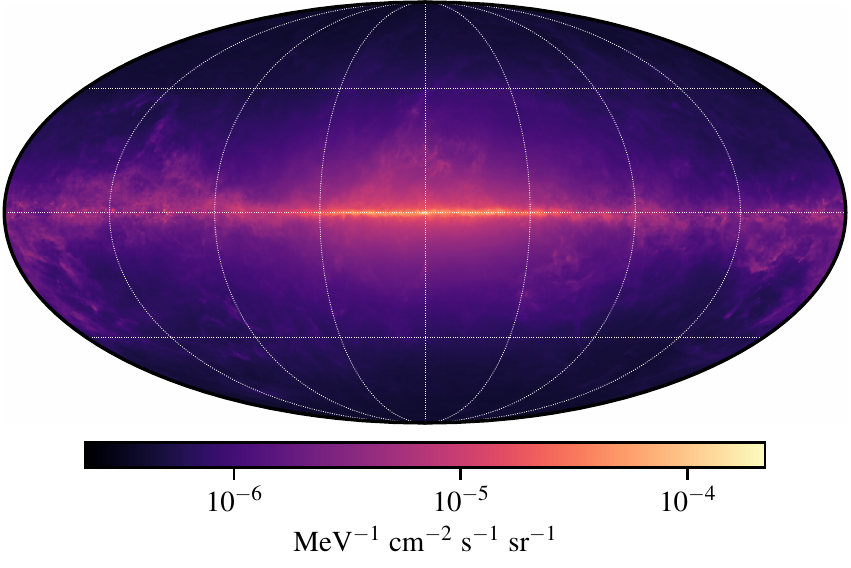}\hfill
\includegraphics[width=0.32\textwidth]{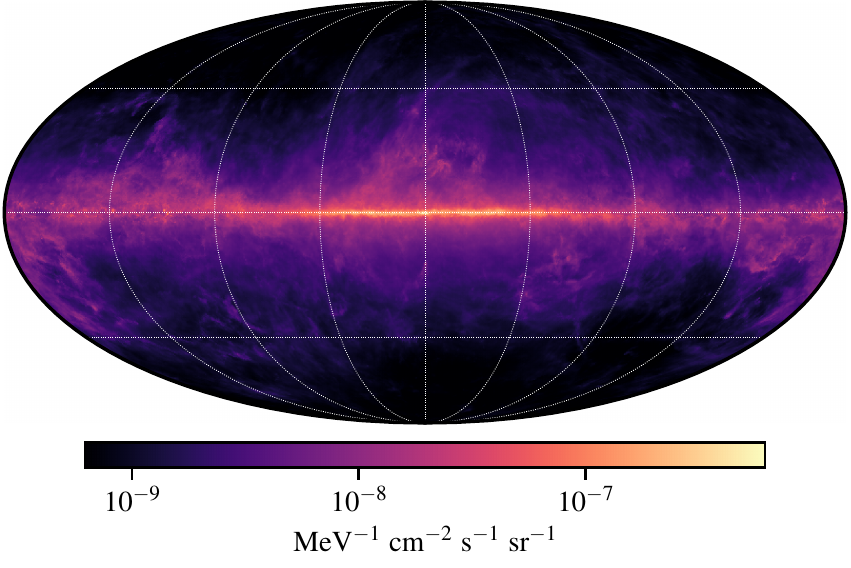}\hfill
\includegraphics[width=0.32\textwidth]{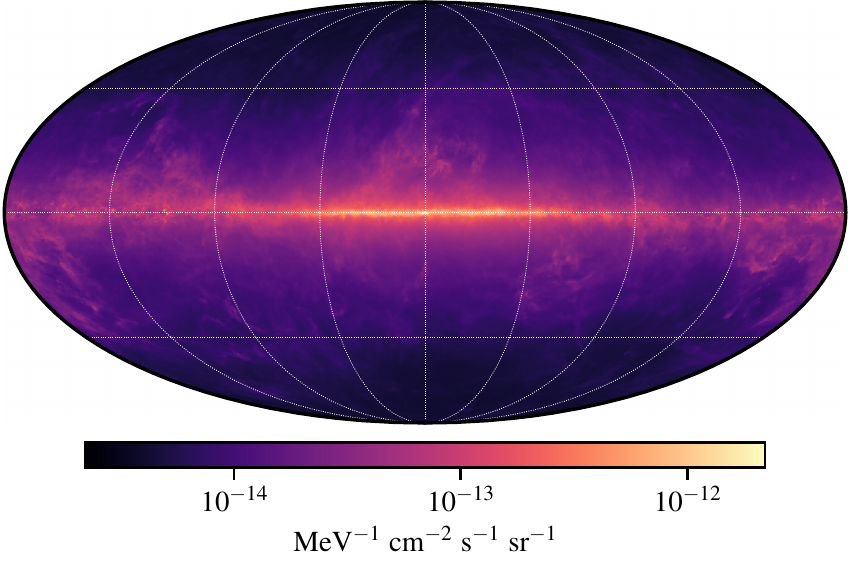}\\
\includegraphics[width=0.32\textwidth]{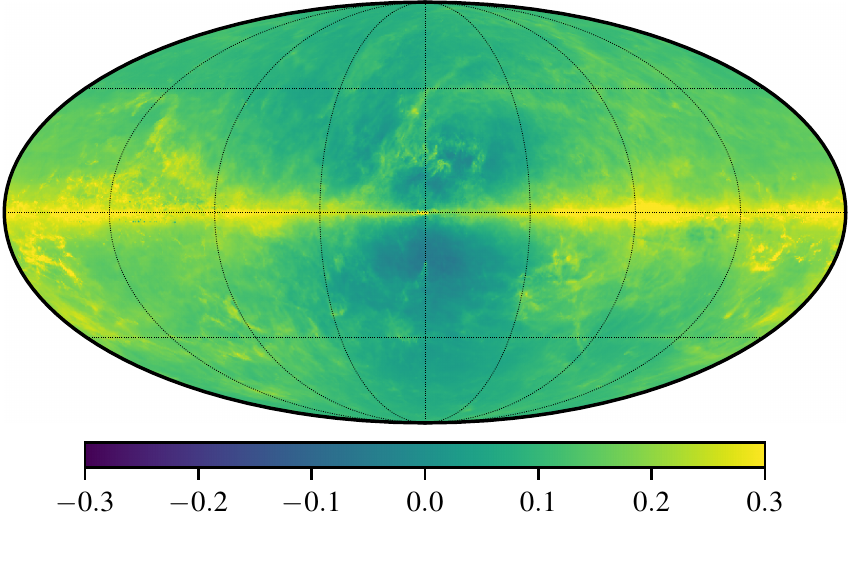}\hfill
\includegraphics[width=0.32\textwidth]{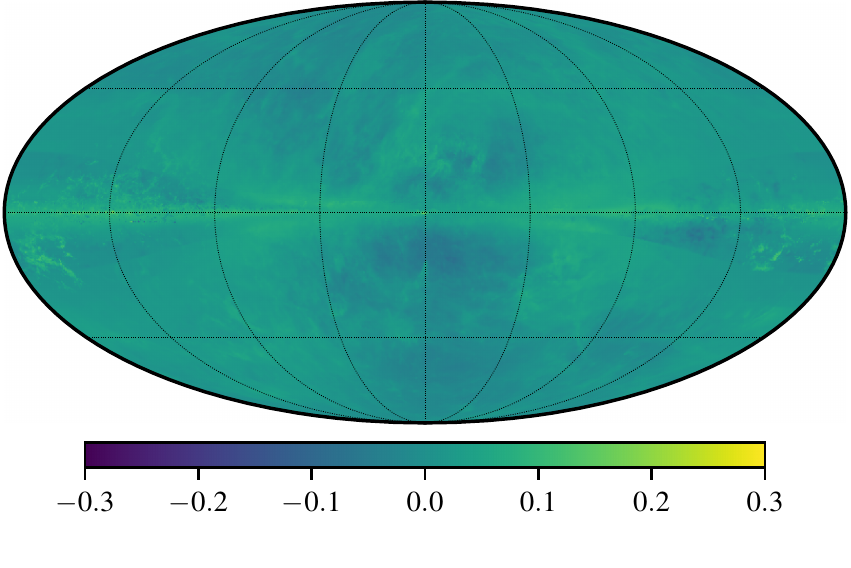}\hfill
\includegraphics[width=0.32\textwidth]{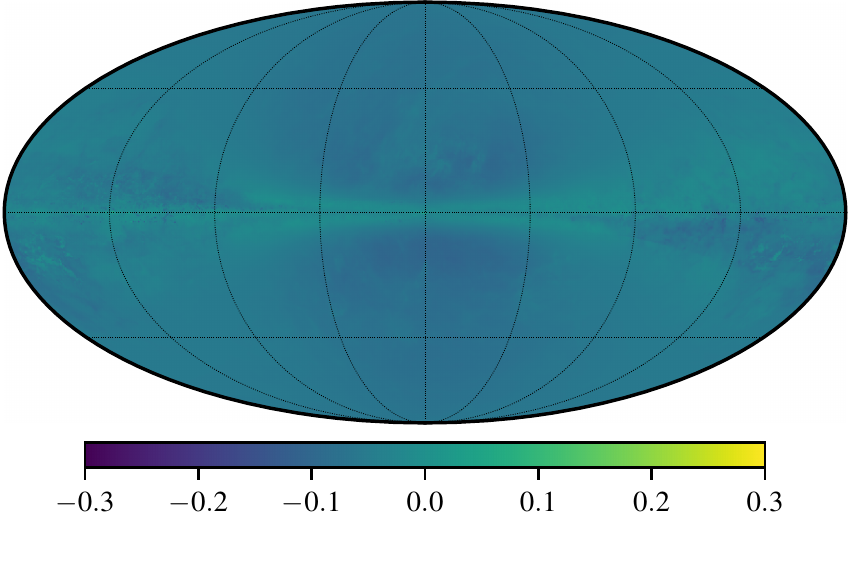}
}
\caption{Top row: total \gray{} intensity ($\pi^0$-decay, bremsstrahlung, and IC) at 30~MeV, 1.0~GeV, and 100~GeV energies (left to right, respectively) for SA0-2D gas reference case. Bottom row: fractional residuals from a comparison with SA0-3D gas \big(\big[SA0-3D -- SA0-2D\big]/SA0-2D\big) at the same energies. The maps are in Galactic coordinates with $(l,b)=(0^\circ,0^\circ)$ at the centre with $l$ increasing to the left. The longitude meridians and latitude parallels have $45^\circ$ spacing.}
\label{fig:0_armsFraction}
\end{figure*}

The total energy density of CRs is dominated by protons with energies of about few GeV that are mostly primary in origin.
The gas distribution only affects the primary CRs by changing the cooling and spallation rates and, therefore, has a minor impact on the total energy density.
Secondary CRs are produced in interactions between primary CRs and interstellar gas, so the gas distribution has a much larger influence on CR secondaries and their spatial densities.
This is illustrated in Figure~\ref{fig:CRsecondaries}, which shows the energy density of secondary positrons in the plane for selected models.
The spiral arm distribution of the gas is clearly visible in the
energy density distribution of the secondary particles, while the
primary CR source distribution has only a relatively minor effect (Figure~\ref{fig:CRdensity}). 

The different drivers for the spatial structure of the primary and secondary CRs result in non-trivial dependence of derived quantities, such as the secondary/primary ratios if the CR source and gas spatial densities are not
aligned.
This is illustrated in the B/C ratio shown in Figure~\ref{fig:CRsecondaries}.
There is a clear reduction in the ratio along the spiral arm pattern of the CR source distribution (black dashed curves), while the ratio is seen to be larger along the spiral arm pattern of the gas distribution (cyan dotted curves).
For the cases where the spiral arm patterns of both distributions aligns the effect of each almost cancels out.
To further illustrate this point, the energy dependences of the B/C ratio at three selected locations in the Galaxy are shown
in Figure~\ref{fig:secondaryRatios}.
The locations are shown in the bottom right panel of Figure~\ref{fig:CRsecondaries} and are chosen to be at the same Galactocentric distance $R_\odot$.
The locations align with a spiral arm in the gas distributions (location L2),
a spiral arm in the CR source distribution (location L3) or both (location L1).
The exact CR and gas distributions can, therefore, have a large effect on the determination of the parameters of propagation models when calculating secondary production and the B/C ratio.
The effect on the pure secondaries, such as secondary positrons shown in
  the bottom row of Figure~\ref{fig:secondaryRatios} is dominated by the gas
  distributions because the
distribution of primary sources have only a small effect.

\begin{figure*}[tb!]
\center{
\includegraphics[width=0.32\textwidth]{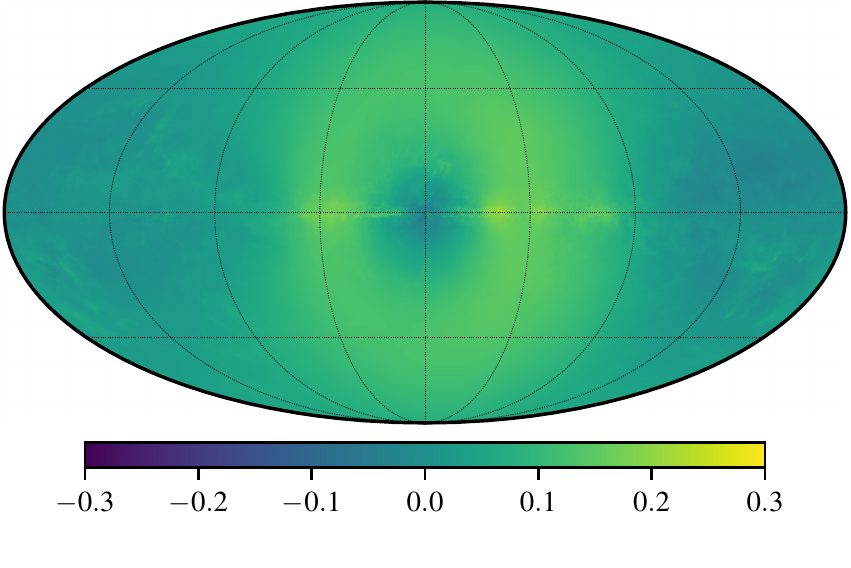}\hfill
\includegraphics[width=0.32\textwidth]{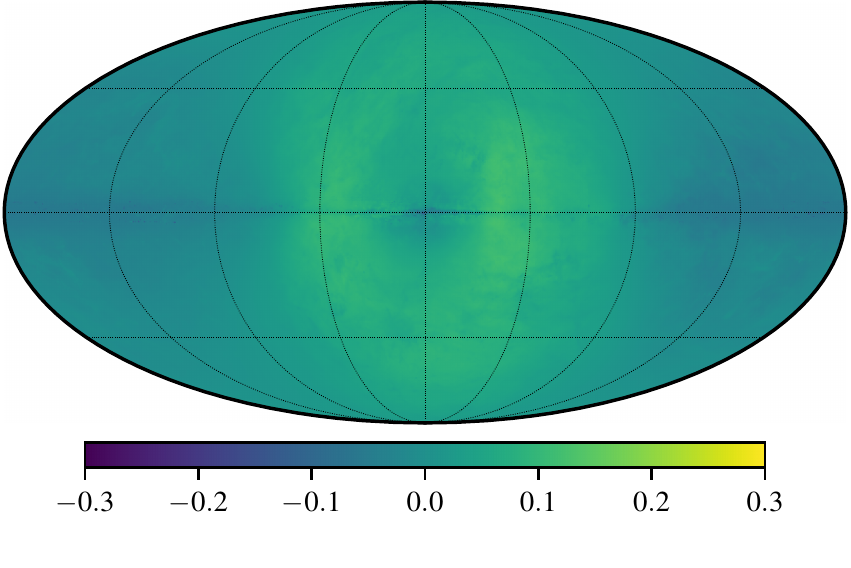}\hfill
\includegraphics[width=0.32\textwidth]{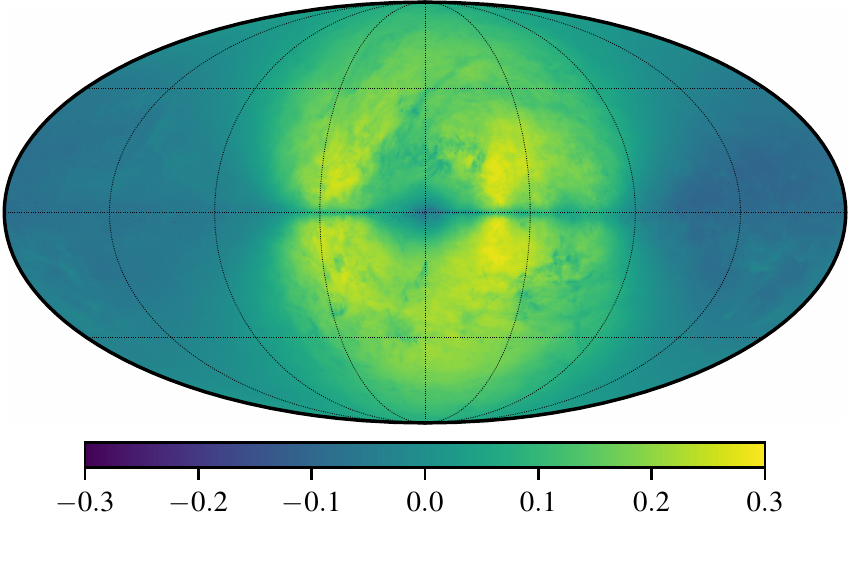}\\
\includegraphics[width=0.32\textwidth]{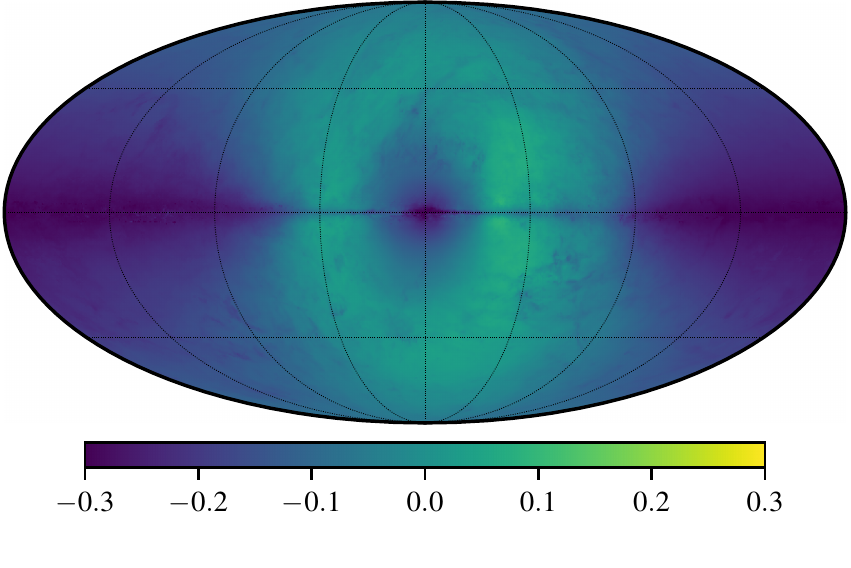}\hfill
\includegraphics[width=0.32\textwidth]{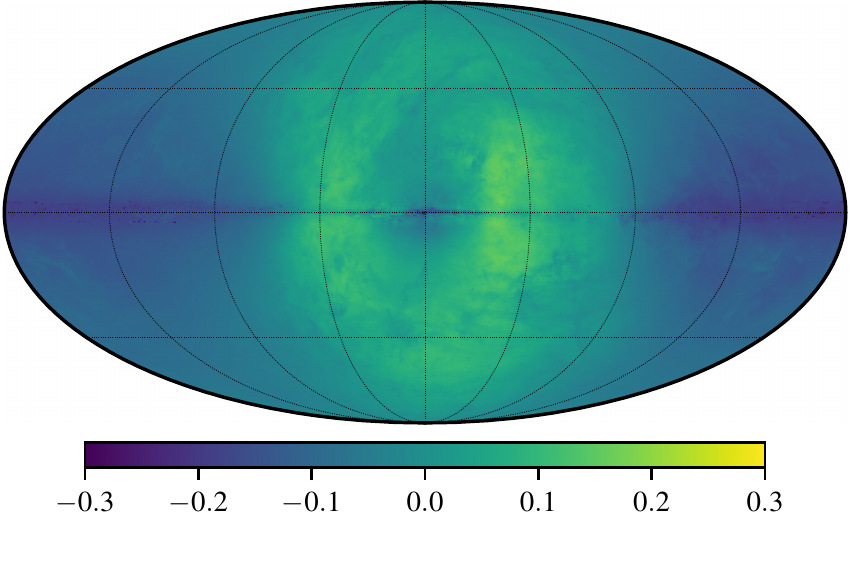}\hfill
\includegraphics[width=0.32\textwidth]{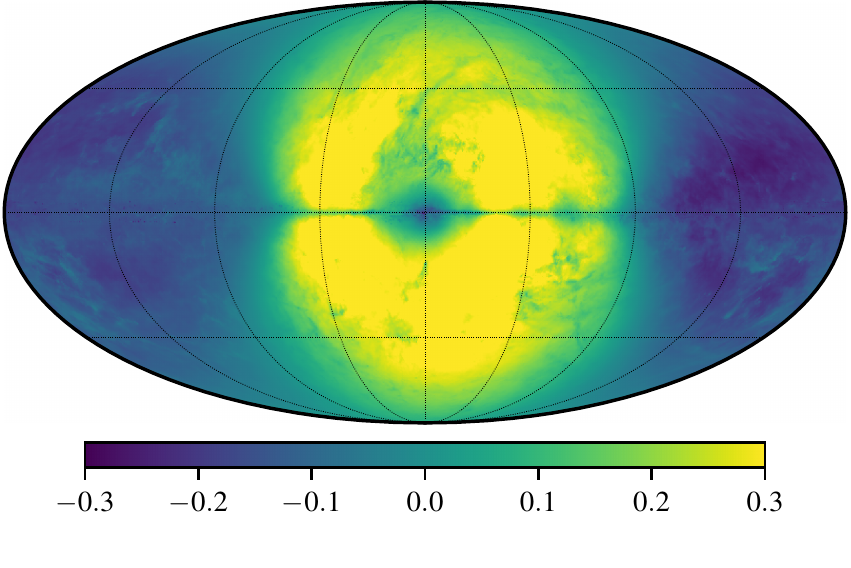}\\
\includegraphics[width=0.32\textwidth]{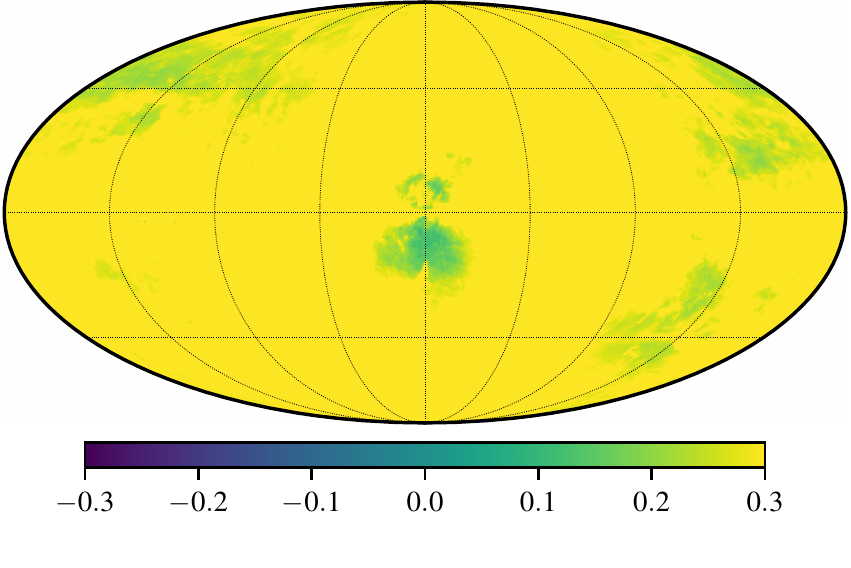}\hfill
\includegraphics[width=0.32\textwidth]{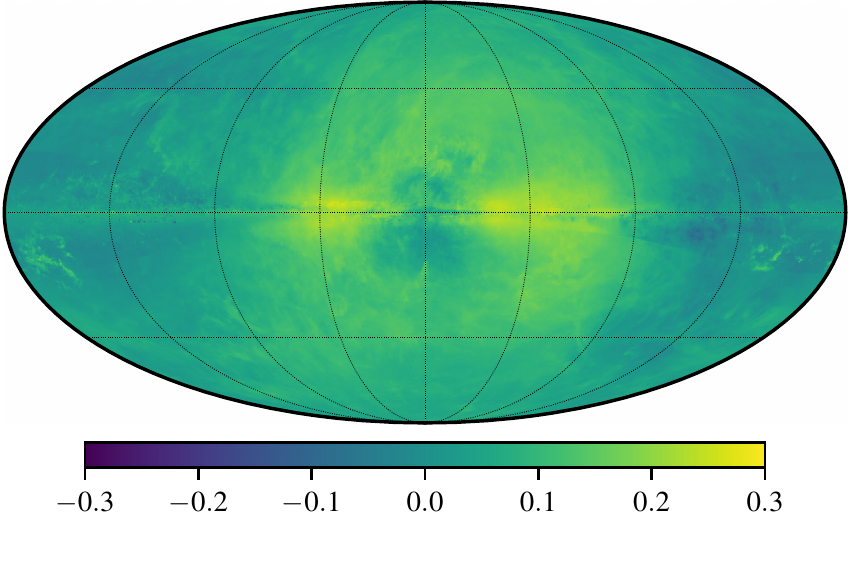}\hfill
\includegraphics[width=0.32\textwidth]{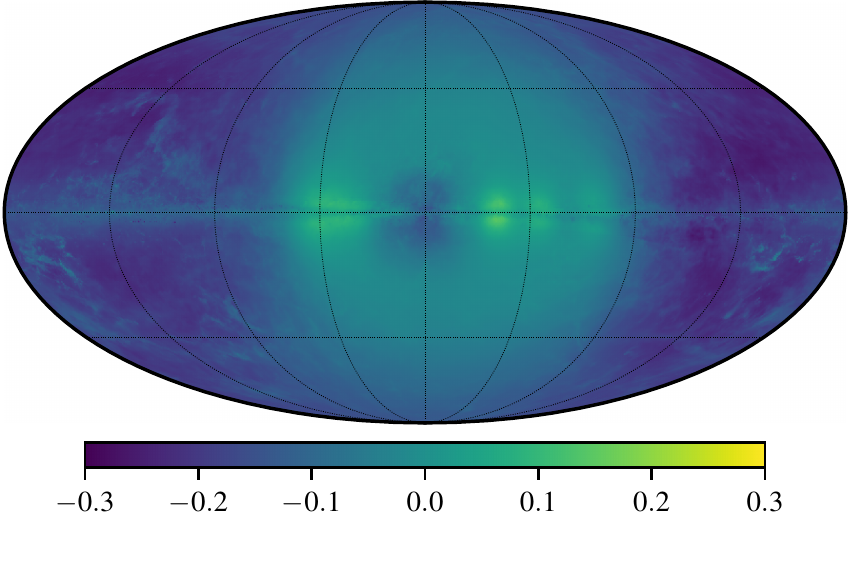}\\
\includegraphics[width=0.32\textwidth]{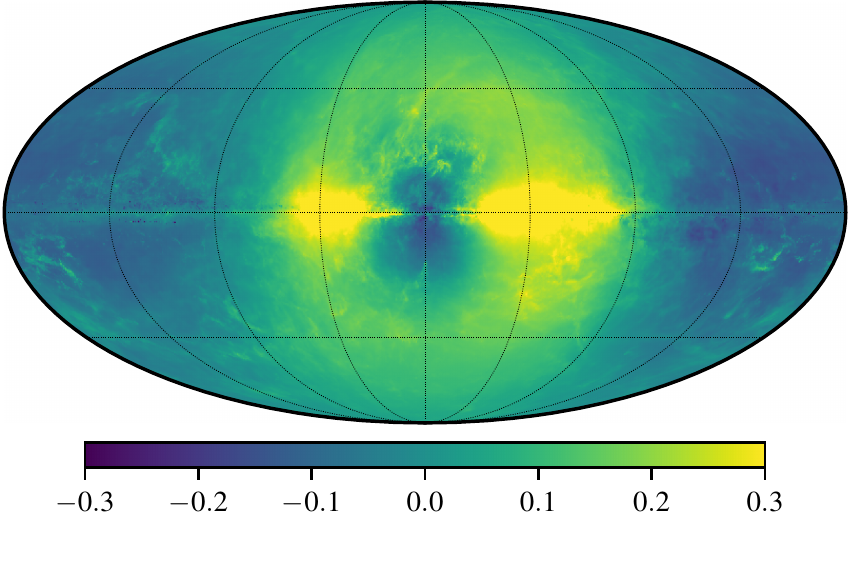}\hfill
\includegraphics[width=0.32\textwidth]{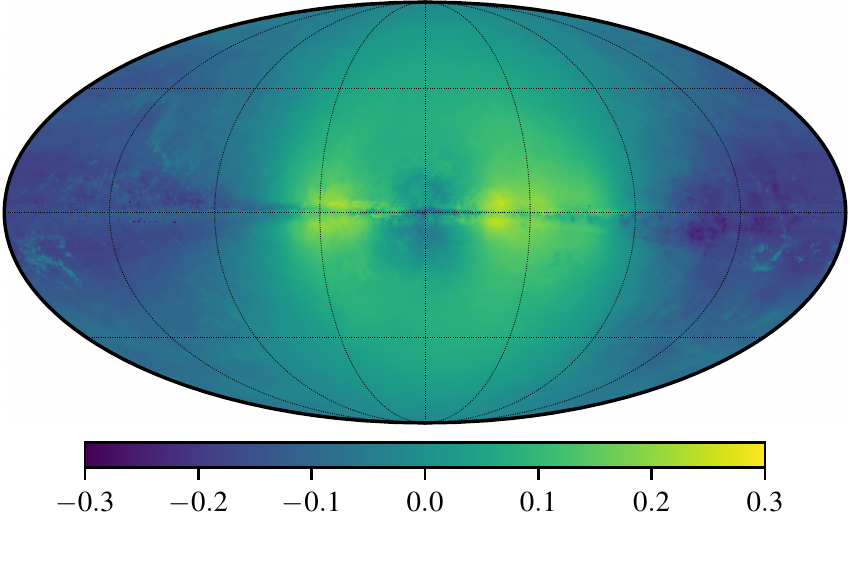}\hfill
\includegraphics[width=0.32\textwidth]{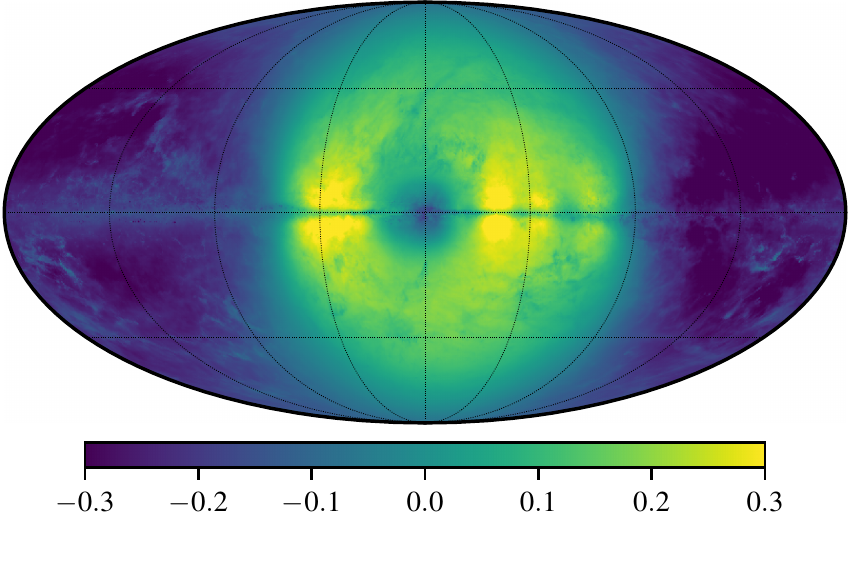}
}
\caption{Fractional residuals for the total \gray{} intensity ($\pi^0$-decay,
Bremsstrahlung, and IC) at 30~MeV, 1.0~GeV, and 100~GeV energies (left to
right, respectively) for SA50-2D gas, SA100-2D gas, SA50-3D gas, and SA100-3D
gas (top to bottom, respectively) compared to SA0-2D gas reference model. The
maps are in Galactic coordinates with $(l,b)=(0^\circ,0^\circ)$ at the center
with $l$ increasing to the left. The longitude meridians and latitude
parallels have $45^\circ$ spacing.}
\label{fig:armsFraction}
\end{figure*}

\subsection{\gray{} maps}

High-energy interstellar emissions are calculated using \GP{} for the SA0,
SA50, and SA100 source density models (Section~\ref{sec:crcalc}), and the
standard 2D gas \citep{AckermannEtAl:2012} and the new 3D gas
model distributions.
The standard 2D ISRF model
\citep{AckermannEtAl:2012} and the annular gas maps from
\citet{AjelloEtAl:2016} are used for all calculations.
Because of the column
density corrections described in Section~\ref{sec:GALPROP}, replacing the 2D
gas distributions in \GP{} with the new 3D gas distributions will only affect
the \gray{} skymaps through the change they have on the CR flux. 

Calculations of the interstellar emission are made with the same spatial and
kinetic energy grids that were used for tuning CR propagation parameters
(Section~\ref{sec:crcalc}).
The \gray{} intensity maps are calculated for the energy range 30~MeV -- 300~GeV using a logarithmic energy grid with 10 bins/decade spacing.
Production of higher energy \gray{s} involves interactions of CRs with energies above several TeV, where the assumption of a steady-state CR injection is less valid due to the stochastic nature of CR sources and fewer sources capable of accelerating particles to VHEs \citep{MoskalenkoEtAl:2001,BernardEtAl:2013}. 
Calculations of pion (and heavier mesons) production and decay are done using a parameterization by \citet{KamaeEtAl:2006}.
Secondary $e^\pm$ produced in hadronic interactions are combined with primary electrons to calculate IC scattering.
Their contribution to the interstellar emissions is important for energies $\lesssim50$~MeV \citep{PorterEtAl:2008,BouchetEtAl:2011}.
All calculations of the IC emission use the anisotropic scattering cross section \citep{2000ApJ...528..357M} that accounts for the full directional intensity distribution of the ISRF model.

The effects the two different gas
distributions have on the total predicted \gray{} intensity are first explored.
 The differences between the model calculations in this paper are of the
  order of few tens of percent, which is much smaller than the variations in
  the calculated sky intensity that are few orders of magnitude, e.g., between
low and high-latitude regions. Such differences between models are most
usefully illustrated as relative differences compared to the reference
case: SA0-2D~gas.
Modifications in the gas distributions used in \GP{} affect the propagation parameters and hence all three components of the interstellar emission: $\pi^0$-decay, Bremsstrahlung, and IC.
Figure~\ref{fig:0_armsFraction} shows a comparison between SA0-2D and SA0-3D model intensities.
The top panels show the total intensity calculated in the SA0-2D model at 30~MeV, 1~GeV, and 100~GeV (left to right), while the bottom panels show the fractional difference between the two models, \big(SA0-3D -- SA0-2D\big)/SA0-2D, for the same energies.
The intensity at 30~MeV is dominated by Bremsstrahlung and IC emission from low-energy electrons with a large contribution from secondary electrons and positrons.
At 1~GeV the intensity is dominated by the $\pi^0$-decay emission where 10~GeV particles contributing the most.
The intensity at 100~GeV is still dominated by $\pi^0$-decay emission over most of the sky with IC emission contributing nearly equally in the inner Galaxy.

The most visible feature of the bottom ratio maps in
Figure~\ref{fig:0_armsFraction} is the positive residuals in the 30~MeV map,
in particular for the local molecular clouds at low latitudes.
This extra
emission is almost entirely caused by increased Bremsstrahlung emission even
though the models are tuned to the same electron data
(Section~\ref{sec:crcalc}).
Because of the smaller number density of the gas
in the 3D model and corresponding change in the propagation parameters, the
local interstellar spectrum of electrons appears softer at low energies than
for the case of the 2D gas model.
Tuning to the same AMS-02 data thus requires a larger modulation potential.
The softer interstellar spectrum of electrons
leads to the enhanced Bremsstrahlung and IC emission compared to the
reference case.
The IC emission is, however, not as strongly affected because
the average energy of the electrons generating most of the emission in this
energy range is higher than for electrons producing Bremsstrahlung.
Higher Bremsstrahlung \gray{} production also enhances the emissions from local gas that is distributed toward high-latitudes, but this is not
visible in the ratio of the total intensity maps because IC emission dominates there.
The effect is greatly reduced for GeV \gray{s}, because Bremsstrahlung becomes subdominant at these energies. 

Other effects visible at all energies are due to the difference in propagation
parameters for the CR source and gas density combinations and the
different spatial distribution of secondary CR particles in the Galaxy.
The flux of secondary CRs in the outer Galaxy is smaller in the model using the 3D gas
distributions compared to the reference case.
The slower diffusion,
however, results
in more primary CR electrons in the outer Galaxy and less in the inner Galaxy
compared to the reference case.
This causes the bluish regions towards the
inner Galaxy at intermediate latitudes where IC dominates over Bremsstrahlung
and $\pi^0$-decay emission.
The enhancement of the IC emission in the outer
Galaxy is not visible in these total ratio maps because
it is subdominant, particularly at 1~GeV.
The decreased emission in the outer
Galaxy, which shows as a sharp drop in the fractional residual maps at the
annular gas map boundaries in the outer Galaxy, is a result of less
Bremsstrahlung and $\pi^0$-decay emission in the outer Galaxy.
Emissions from local clouds dominate the intensity in the outer Galactic
plane, which explains why the ratio is slightly above 1 for the local clouds.
The spiral arm
structure of the 3D gas distributions is not visible in the fractional ratio
maps because they affect the total \gray{} intensity by only a few percent,
which is smaller than the effect caused by the change in propagation.
Even though the effects of the spiral arm features are small in the
total intensity the magnitude of the difference can be up to 15\% for individual Galactocentric annular maps (not shown).

Comparisons of models with CR sources in the spiral arms, SA50 and SA100, are
shown in Figure~\ref{fig:armsFraction} as fractional residuals against the
reference model.
Shown are residuals for both the 2D and 3D gas distributions.
The spiral arm contribution in the CR source distribution falls off more
quickly with Galactic radius than the disk contribution.
This causes reduced
emission in the outer Galaxy for models with a higher fraction of sources in
the spiral arms.
The same effect also causes an increased flux of CRs in the
inner Galaxy that produces an increased intensity towards the inner Galaxy,
but the lack of sources at $R < 3$~kpc in the spiral arm component results in
a reduced intensity near the GC.
This combination produces the doughnut-like
shape in the residuals towards the inner Galaxy.
The effect is enhanced for
the CR electrons because they lose energy more quickly than nuclei and,
therefore, do not travel very far from their sources.
The enhancements around the spiral arm tangents are, therefore, mostly visible in Bremsstrahlung and IC emission components.
These can be seen in 30 MeV and 100 GeV fractional residual maps, where one of these components or both are bright enough.
Correspondingly, the tangents are least visible in the fractional residuals at
1~GeV where the $\pi^0$-decay component dominates. 

The increased brightness of the 30~MeV map for the SA50-3D gas model is
due to the larger value needed for the AMS-02 modulation potential to match
the CR data (see Table~\ref{tab:CRparameters}).
As discussed before there is considerable degeneracy between the determination of the heliospheric modulation and the low-energy CR intensities.
Combined with non-linear CR propagation models and numerical
uncertainties, this can lead to the minimizer having difficulties in finding
the true best-fit parameters when fitting to CR data.
This may explain why the values for the modulation potential differ so significantly between the three models using the 3D gas distribution.
\fermilat\ data can be used to further constrain the spectrum of CR electrons, but such analysis requires modeling outside the scope of this paper and is deferred to future work.

The effects of variations of the CR source distribution on the \gray{}
intensity maps are fairly similar in cases of 3D and 2D gas distributions.
There is a reduction in the intensity in the outer Galaxy and toward the GC
while the intensity in the inner Galaxy increases.
The details are, however, somewhat different.
The slower diffusion combined with the same electron cooling
rate leads to more CR electrons near their sources and fewer far from the
plane.
The doughnut like excess towards the inner
Galaxy is thus more asymmetric with higher intensity near the plane.
This is most visible at 100~GeV where the increased intensity for latitudes $|b| \gtrsim 30^\circ$ is suppressed compared to the same source models with the 2D gas distribution.
This effect also enhances and sharpens the spiral arm tangent features
for both IC and Bremsstrahlung.
The CR nuclei are not as strongly affected because the
change in the gas number density and hence the cooling and fragmentation rates
compensate to some extent for the change in the diffusion coefficient.

\section{Discussion}

The 3D gas density models derived in this work employ only a few spatial distribution components, which are able to account for the main features of the interstellar gas with relatively few parameters.
This enables the model fitting to be made within a reasonable time frame.
Among the elements held constant for the tuning of the gas distribution models is the gas rotation field, which defines the conversion between distance and velocity.
To test the effect of varying the rotation field a fit to the gas data was performed using a model with the rotation curve of \citet{Clemens:1985}, but other components were kept the same.
Using this rotation curve resulted in an overall worse fit of the data, as determined by the log-likelihood.
Considerably more overlap between spiral arms was evident, and the features in the longitude--velocity diagram of the model did not match those in the data as well as the best-fit model determined above.
The smooth disk also contained a larger fraction of the total mass of the model. The assumed velocity field is, therefore, very important for accurate determination of a 3D spatial density model for the interstellar gas.

\begin{figure}[t]
 \centering
 \includegraphics[width=0.5\textwidth]{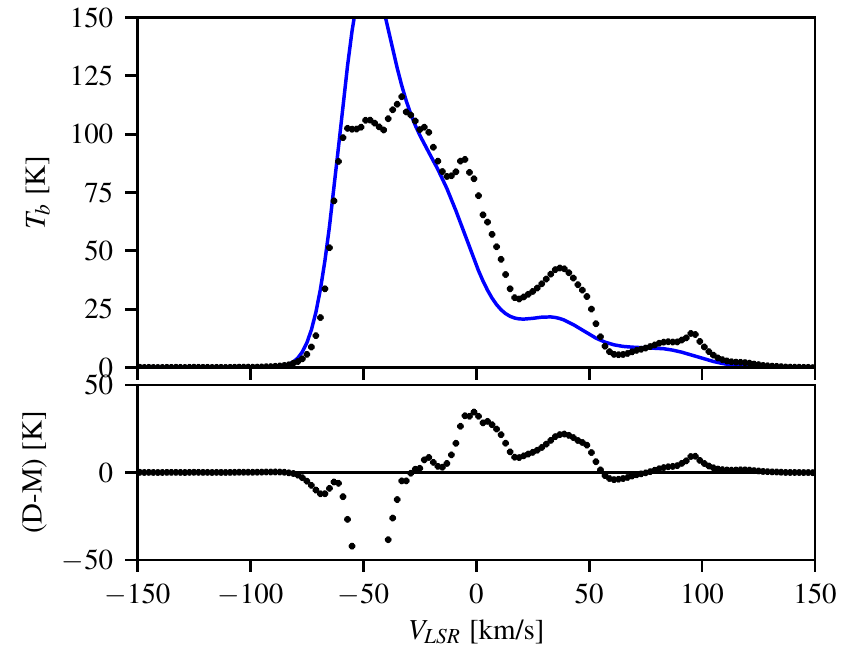}
 \caption{The \hi\ velocity spectrum in the direction of $(l,b)=(-50^\circ,0^\circ)$. Data are shown as black points with error bars smaller than the point size. The blue curve shows the model calculations. The residuals are shown at the bottom panel.}
 \label{fig:velocitySpectra50}
\end{figure}

Inclusion of a detailed description for the non-cylindrical streaming motions into the models would add another level of complexity to the analysis and was not attempted in the present work. However, to test their effect a simplified sinusoidally varying radial component 
\begin{equation}
  v_r(\Theta) = v_{r,0} \sin(\Theta - \theta_{r,0})
  \label{eq:vr}
\end{equation}
\noindent
was added to the cylindrical rotation with $v_{r,0}$ and $\theta_{r,0}$ as free parameters.
The same model fitting procedure was then followed as for the \hi\ and CO models described above    
with all components included.
The likelihood improved with this additional component, but not by enough to warrant its inclusion into the final models.
The improvement in the likelihood came almost entirely from a somewhat closer match to the data for the \hi\ model.
The total mass for the atomic and molecular gas density models increases with this modification, which is expected for an improved fit.

The total gas mass in the \hi\ model increases to $6.0\times 10^9$~M$_\odot$ with almost all increase due to the spiral arms.
The radial number density distributions of the disk and arms also become flatter, being nearly constant from 2 kpc to 15 kpc from the GC and varying only by $\sim$25\% over the entire range.
The gas mass in the CO model slightly increases to $0.74\times 10^9$~M$_\odot$, where most of the increase is due to the bulge/bar component.
The radial density distribution of the CO disk/arm component is also flatter, having about 20\% of the disk mass in the outer Galaxy compared to 5\% without the velocity modifications. 

The mophology of the spiral arms also changes: the pitch angles for all
arms fall in between $12.0^\circ$ and $13.5^\circ$ with all starting points in
the range 3.1--3.6 kpc.
The spiral arms are thus more symmetric in their shapes.
The arm densities for the CO model are still very asymmetric with both arm 1 and arm 2 set to zero by the fit to the CO data.
In particular, arm 2 that has been identified with Sagittarius and
Carina in the model still does not match the features in the CO data.
The flaring parameter, $r_z$, increases and is now in better agreement with that
determined by \citet{KalberlaDedes:2008}.
The warp changes slightly with all the warp-related parameters increasing more slowly with radius without changing its azimuthal dependence.

The resulting velocity field is broadly consistent with that found by
\citet{TchernyshyovPeek:2017} and shown in their Figure~5.
There are negative velocity corrections close to the GC and in the outer
Galaxy for longitudes larger than $180^\circ$, while the corrections are positive in the outer Galaxy for longitudes smaller than $180^\circ$.
The only large-scale feature they found that is not reproduced is the positive peculiar velocity around longitude $\sim45^\circ$.
The final value for $v_{r,0}$ was around 28~km~s$^{-1}$, so the magnitude of the variations is also similar to that obtained by \citet{TchernyshyovPeek:2017}.
This shows that streaming motions are important to consider, but also that these simple modifications to the velocity field in combination with the geometrical model derived in this paper do not qualitatively change the final results.
In particular, the effect on the CR propagation parameters is only minor.

The neutral hydrogen in the interstellar gas is not uniform.
It is composed of, at least, three different phases that can be separated by their temperature: the stable cold, the stable warm and the unstable warm \citep{VerschuurMagnani:1994}.
Observations have shown that each of these phases contain about one third of the total mass of the neutral hydrogen \citep{HeilesTroland:2003}.
Different temperatures of these phases result in widely different broadening of the line emission profiles, ranging from few km~s$^{-1}$ for the cold phase up to few tens of km~s$^{-1}$ for the warm stable phase. 

The average density of the gas in these phases is also different with the cold gas generally being about an order of magnitude denser than the warm gas.
The single temperature analysis employed in this work uses the optically thin approximation that is appropriate only for the warm neutral medium and cannot be expected to capture all density structure of the neutral hydrogen.
In particular, the cold component should exhibit smaller average line broadening than employed in this work.
The spiral arm components are the densest parts of the \hi{} model and are also likely to have a larger fraction of the cold medium than the more diffuse disk component, therefore, their line profiles are narrower than the disk line profiles. 

Colder gas is also more likely to be optically thick.
The gas number density
in the structures consisting of the cold phase is thus under-predicted in this
work because of both, the model inability to reproduce the observed
narrow emission lines, and the optically thin assumption that over-predicts the line emissivity per hydrogen atom.
An improved treatment that accounts for the multi-component neutral medium is necessary to simultaneously account for the cold and warm phases.
The results of the current paper should be considered as providing an approximate lower bound to the true gas number density in the Galaxy.

Even though the models developed in this paper generally under-predict the data, there are a few exceptions.
An example can be found in the \hi\ longitude profiles shown in Figure~\ref{fig:longitudeProfiles}.
The model over-prediction toward the GC can be seen, though the number
density in the \hi{} model becomes nearly 0 at the GC (see
Table~\ref{tab:diskParameters} and Figure~\ref{fig:gasMaps}).
The over-prediction is likely a consequence of using the optically thin approximation in Eq.~(\ref{eq:HItransport}).
To test this hypothesis, the analysis was redone assuming a constant
$T_S=150$~K
throughout the Galaxy.
This results in a lower brightness
temperature toward the inner Galaxy despite the gas number density being
non-zero and the total \hi{} mass being $\sim10$\% higher than determined using the optically thin assumption.
The reduced brightness is the result of absorption from foreground
gas that has higher opacity in the model with lower $T_S$.
The likelihood of the model with the
lower $T_S$ value is, however, significantly lower.
This simple modification
may help data agreement in the direction toward the GC, but other regions then are not fitted very well.

The optically thin assumption may also affect the model-data agreement in other regions.
In particular, near the Crux-Centaurus arm tangent at $l \approx -50^\circ$, the model significantly over-predicts the data.
This is shown in the velocity spectra in the LOS towards
$(l,b)=(-50^\circ,0^\circ)$ in Figure~\ref{fig:velocitySpectra50}.
Using $T_S \sim 110$ K, the peak of the $T_S$ distribution estimated by \citet{StrasserTaylor:2004}, would reduce the model prediction to agree well with the data in that direction.
However, the assumption of a single value for $T_S$ over the entire Galaxy
would unlikely result in a global improvement in the model-data agreement
because of the non-linearity of the optical depth correction, Eq.~(\ref{eq:HItransport}).
Instead, modeling $T_S$ variations throughout the Galaxy that takes into account observations of \hi{} in absorption, where possible, may help.

The modifications of the \gray{} maps due to the implementation of the new
3D gas distributions can almost be entirely attributed to changes in the calculated CR flux.
This is because the final maps are scaled with column densities determined from the line emission surveys for each LOS.
Because of large uncertainties in the properties of the CR source distributions these changes may be compensated for, at least partially, by variations in the CR source distribution.
This work has shown that changing the CR source distribution has a larger effect on the \gray{} maps than changing the gas distributions used by the propagation codes.
The correlation between the gas distribution and CR source distribution does not mean, however, that using a more realistic gas distribution is unnecessary.
On the contrary using the best available gas distribution is vital if attempting to use \gray{} observations to constrain the properties of CR sources and propagation. 

The most popular method for determining the CR source distribution from \gray{} data is to use the annular maps created from the line emission surveys and assume the CRs illuminate the gas uniformly.
This is a useful first approximation, but the calculations made in this paper show that implementation of the 3D distributions of CR sources and components of the interstellar medium can result in variations of tens of percent within a single annular map, which is far greater than the statistical uncertainty of current data that is less than one percent.
Also, the annular maps may provide excellent resolution and details on the sky, but they suffer from poor distance resolution and near-far distance ambiguities. The 3D gas models are, therefore, necessary to accurately weight the CR flux within the region covered by the annular map.
This is even more important if 3D source density distributions are used. 

The SA0--2D model is based on the same assumption as the $^{\rm S}$L$^{\rm Z}6^{\rm R}20^{\rm T}150^{\rm C}5$ model of \citet{AckermannEtAl:2012} and the Pulsars model of \citet{AjelloEtAl:2016}.
It is interesting to compare the fractional residual maps in
Figure~\ref{fig:0_armsFraction} and Figure~\ref{fig:armsFraction} to the fractional residual maps presented in these two earlier studies.
In Figure~7 of \citet{AckermannEtAl:2012} there is a general trend of negative residuals for the latitudes ranges $5^\circ < |b| < 15^\circ$ that corresponds to the reduction in IC emission at these latitudes when using the 3D gas distribution.
This is also in agreement with the scaling factor for the ISRF in that work being less than 1 in their inner Galaxy region.
There is also indication of asymmetry in the residuals along the plane with positive residuals at $l\sim45^\circ$ and negative residuals at $l\sim-45^\circ$.
A similar pattern is also seen the residuals in Figure~2 of \citet{AjelloEtAl:2016}.
While none of the models in this work show this exact behavior it is not difficult to imagine modifications to the CR source distribution that could naturally account for such model asymmetries.

\section{Summary}

New 3D models for the large-scale distributions of atomic and molecular gas in
the Galaxy have been developed.
They are based on a combination of a limited
number of geometric components with smoothly varying number density that
enables the description of the observed large-scale structure: the warped
disk, the central bulge/bar, and the 4 major spiral arms. The parameters of
model components have been tuned to match the LAB 21-cm line emission survey
\citep{KalberlaEtAl:2005} and the CfA composite CO survey
\citep{DameEtAl:2001}.
The best-fit parameters and resulting models agree very
well with previous work, but it is the first time that they are presented
together within a single consistent framework.
To study the effects
of variations in the gas and source distributions on CR propagation
parameters they are incorporated into the GALPROP CR
propagation code.
The new gas density models significantly affect the best-fit
values of CR propagation parameters determined from the fits to local CR data.
The parameter values depend on both the CR source density distribution
and the gas density models and will have to be re-evaluated as better models
are constructed.

The effects that three different CR source density
models with different injection rates in the smooth disk and spiral arms
components have on the \gray{} intensity skymaps have also been investigated.
The combined 3D CR source and gas density models produce non-trivial features in the large-scale \gray{} residual sky maps that can be compared with those obtained from the prior analysis of the interstellar \gray{} emissions observed by the \fermilat.
Elements of the 3D models provide an explanation for residual features previously obtained from analysis of \gray{} data using the 2D axisymmetric models. 

\acknowledgements
\GP{} development is partially funded via NASA grant NNX17AB48G. Some of the results in this paper have been derived using the  HEALPix~\citep{GorskiEtAl:2005} package.

\appendix
\section{\GG{} description}
\label{app:GALGAS}

The \GG\ code is designed to calculate line-emission profiles from the Galaxy
for user-provided gas density and velocity distributions.
Efficient evaluation
of a likelihood function many times is necessary for the model-data fits. The
code is written in C++ and is parallelized using
OpenMP\footnote{\href{http://www.openmp.org}{http://www.openmp.org}} for CPUs and
OpenCL\footnote{\href{https://www.khronos.org/opencl/}{https://www.khronos.org/opencl/}}
for execution on GPUs (the code operates using both if available on a given
system). The code calculates the line-emission profile on a HEALPix grid.
The emission of each pixel is uniform and its value determined from the
LOS going through the center of the pixel.
Each LOS
profile is binned in the same way linearly in velocity.
For each pixel the
code first calculates the transformation from distance to velocity for a given
velocity field $\boldsymbol{v}(X,Y,Z)$ by calculating the difference in
velocity between the location of the Sun and the location along the LOS,
projected onto the LOS.
The code steps through the LOS beginning from the Sun
and finding the distance points corresponding to the boundaries of the velocity bins.
A coarse search is
initially performed using a step size based on 
the gradient of the projected velocity, then the distances are further refined
using a bisection search.
A maximum of 20 bisection steps are taken for each point.
More efficient search methods were found to be less stable than the
bisection method and, therefore, not considered.
The code currently steps along the LOS until it reaches a distance of 100~kpc.
Even though each LOS is independent, this part of the code executes only on the CPU because the branching for the current algorithm is inefficient for execution on the GPU.
The results from the transformation calculations are cached internally and
only recalculated as needed if the parameters of the velocity field change.

Once the transformation from velocity to distance has been established, the
actual LOS integration is performed on the GPU. Each distance bin is
independently integrated using 61 point Gauss-Konrod integration rule. The
integrator does not handle radiative transfer, so each bin is assumed to have
constant average optical depth for absorption. For CO, the line emission is
assumed to be optically thin and therefore linearly related to the density,
while for \hi\ Eq.~(\ref{eq:HItransport}) is used to estimate the line
intensity. The $T_S$ is calculated as the average of a given $T_S(X,Y,Z)$
distribution over the distance bin. While the integration over the LOS is
performed on the GPU, the transformation from the distance bin grid to the
velocity grid is calculated on the CPU. The model is then smoothed in velocity
with a Gaussian kernel to account for thermal and turbulent motions of the
gas. The width of the kernel is parameterized by velocity and then internally
binned to match the specified velocity grid. The actual convolution with the
kernel is performed on the CPU. Profiling of the code revealed that performing
this on the GPU would result in small gains only.

The model-data comparison is done using a student-t likelihood with a user
specified degrees of freedom. The input data can be either a HEALPix cube or a
CAR projected cube. The code automatically rebins the CAR cubes into the
HEALPix format and the user specified velocity binning. It can also handle
data that are split into several longitude bins, such as the LAB split data
format. Simple filters can be specified as rectangular regions in longitude,
latitude, and velocity. These are useful to filter out high-velocity data and
Local Group galaxies from the \hi\ data. The filtered and rebinned data can be
cached for future use so rebinning does not happen each time a new fit is
performed. The likelihood calculation is made on the CPU (execution on the GPU
for this part does not increase the performance of the code significantly).

The model is defined using an XML based model format. The density models use
the galstruct library from the {\it GALTOOLSLIB} package distributed with the
\GP{} code. Each density model is composed of different components or building
blocks that is each derived from a base C++ class in the galstruct library.
Adding a new component simply requires writing a new class that defines the
functional form and its parameters.
Once the class is registered with an identification string in
the code it can be used in the XML input file and given a unique name. The velocity field model is
based on a similar concept and is composed of different components as well.
The initial values and boundaries of the fitted parameters are specified in
the XML file. Upper and lower boundaries are set independently and they can be
turned off completely. Parameters can also be fixed to their initial values.
All parameters have a name that is determined from the given component name
and the default name of the parameter.
This way the same component (e.g., a spiral arm) can be included many times in the model with each instance
having its own individual parameters.
This default name of parameters can be overridden with a special name
tag. By assigning the same name to two parameters they become linked in the
fitting procedure and take the same value.

Once the fitting procedure is done and the best-fit parameters have been
derived, the code outputs the model configurations as XML files with updated
values. The code also outputs the parameter values along with their
uncertainties as a text file. By default, the code stores the best-fit model
and the data-model residual as HEALPix cubes. The code can also calculate the
surface density, the first moment of the vertical density distribution
(Eq.~[\ref{eq:meanHeight}]), the vertical thickness of the density
distribution using the definition from \citet{LevineEtAl:2006}, and the
longitude-velocity diagram. The code is configured with a simple text based
parameter file where all the input and options are defined.
\end{document}